\documentclass[fleqn,usenatbib]{mnras}

\usepackage[T1]{fontenc}

\DeclareRobustCommand{\VAN}[3]{#2}
\let\VANthebibliography\thebibliography
\def\thebibliography{\DeclareRobustCommand{\VAN}[3]{##3}\VANthebibliography}

\usepackage{graphicx}	
\usepackage{amsmath}	
\usepackage{amssymb}	
\usepackage{xcolor}

\usepackage{booktabs} 

\newcommand{\beq}{\begin{equation}}
\newcommand{\eeq}{\end{equation}}
\newcommand{\beqn}{\begin{equation*}}
\newcommand{\eeqn}{\end{equation*}}
\newcommand{\mdotsink}{\dot{m}_{\rm sink}}
\newcommand{\rcirc}{r_{\rm circ}}
\newcommand{\rsink}{r_{\rm sink}}
\newcommand{\machr}{\mathcal{M}_r}
\newcommand{\avercirc}{\overline{r}_{\rm circ}}
\newcommand{\risco}{r_{\rm isco}}
\newcommand{\jisco}{j_{\rm isco}}
\newcommand{\jrand}{j_{\rm rand}}
\newcommand{\Mhalf}{M{\small-half} }
\newcommand{\Mhalftp}{M{\small-half-tp} }
\newcommand{\Shalftp}{S{\small-half-tp} }
\newcommand{\Mhalfns}{M{\small-half}}

\newcommand{\Shalftpns}{S{\small-half-tp}}

\newcommand{\ath}{{\sc athena++} }
\newcommand{\mesa}{{\sc mesa} }
\newcommand{\athns}{{\sc athena++}}
\newcommand{\mesans}{{\sc mesa}}
\newcommand{\msun}{M_\odot}
\newcommand{\rsun}{R_\odot}

\newcommand{\Lpl}{L_{\rm pl}}
\newcommand{\tpl}{t_{\rm pl}}

\definecolor{mypink}{HTML}{FF376A}
\definecolor{mygreen}{HTML}{32CD32}
\definecolor{myblue}{HTML}{9932CC}

\title[Explosions from convective supergiant collapse]{Numerical simulations of the random angular momentum in convection II: delayed explosions of red supergiants following ``failed'' supernovae}

\author[A. Antoni and E. Quataert]{
Andrea Antoni$^{1}$\thanks{E-mail: aantoni@berkeley.edu}\href{https://orcid.org/0000-0003-3062-4773}{\includegraphics[width=9pt]{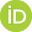}}
and Eliot Quataert$^{1,2}$\href{https://orcid.org/0000-0001-9185-5044}{\includegraphics[width=9pt]{ORCID32}}
\\
$^{1}$Astronomy Department and Theoretical Astrophysics Center, University of California, Berkeley, CA 94720, USA\\
$^{2}$Department of Astrophysical Sciences, Princeton University, Princeton, NJ 08544, USA\\
}

\date{Accepted XXX. Received YYY; in original form ZZZ}

\pubyear{2023}

\begin{document}
\label{firstpage}
\pagerange{\pageref{firstpage}--\pageref{lastpage}}
\maketitle

\begin{abstract}
When collapse of the iron core in a massive red or yellow supergiant does not lead to an energetic supernova, a significant fraction of the convective hydrogen envelope will fall in towards the black hole formed from the collapsing core. The random velocity field in the convective envelope results in finite specific angular momentum in each infalling shell. Using 3D hydrodynamical simulations, we follow the infall of this material to small radii, resolving the circularization radii of the flow.  We show that infall of the convective envelope leads to nearly complete envelope ejection in a $\gtrsim$ 10$^{48}$ erg explosion with outflow speeds of $\gtrsim$ 200 km/s.  The light curve of such an explosion would show a characteristic, red plateau as the ejecta cools and a hydrogen recombination front recedes through the expanding ejecta. Adopting supernova IIp scalings, the event would have a plateau luminosity of $\gtrsim$ 10$^{40}$ erg/s and a duration of several hundreds of days.  These events would appear quite similar to luminous red novae with red or yellow supergiant progenitors; some luminous red novae may, in fact, be signposts of black hole formation. The mechanism studied here produces more energetic explosions than the weak shock generated from radiation of neutrino energy during the proto-neutron star phase. Because we cannot simulate all the way to the horizon, our results are likely lower limits on the energy and luminosity of transients produced during the collapse of a red or yellow supergiant to form a black hole. 
\end{abstract}

\begin{keywords}
black hole physics -- convection -- stars: massive -- supernovae: general.
\end{keywords}



\section{Introduction}
\label{sec:intro}
Theory and observation have unambiguously associated Type IIp supernovae (SNe) with the successful explosions of red supergiants (RSGs) following the collapse of the star's iron core.  Indeed, SNe IIp are the most common type of SNe and RSGs are the most common pre-SN stars expected from single stellar evolution. However, not all RSGs necessarily explode as SN IIp.  It is feasible that some fraction of massive-stellar deaths do not lead to a neutrino-driven SN explosion of the star \citep{2008ApJ...679..639Z,2012ApJ...757...69U,2016ApJ...821...38S,2020ApJ...890..127C,2020MNRAS.491.2715B,2023ApJ...949...17B}.  Instead, the explosion mechanism may fail and a large fraction of the star will fall in towards the black hole (BH) unavoidably formed from the collapsing core.  

In the case of such BH formation in apparently ``failed'' SNe (FSNe), there are nevertheless a few mechanisms that can generate an observable transient.  One way is by the formation of a weak shock due to radiation of neutrinos from the core during the temporary proto-neutron star phase  \citep{1980Ap&SS..69..115N,2013ApJ...769..109L,2018MNRAS.476.2366F,2021ApJ...911....6I,2022arXiv220915064S}. This can lead to a $\sim$months long red transient in the case of RSG progenitors (discussed in more detail in Sec.~\ref{sec:lovegrove}). Another way to generate an explosion after a FSN is via accretion onto the newly-formed BH.  It is well-known that collapse of rapidly rotating stars can generate outflows \citep{2019arXiv190404835B,2020ApJ...901L..24M} with rapidly-rotating Wolf Rayet stars likely powering long gamma-ray bursts \citep{1993ApJ...405..273W,1999ApJ...524..262M,2022MNRAS.510.4962G}; more extended, rotating supergiants may be the progenitors of ultra-long gamma-ray transients \citep*{2012ApJ...752...32W,2012MNRAS.419L...1Q,2018ApJ...859...48P}.   

The cores of RSGs are thought to be slowly rotating at the end of their lives \citep{2012ApJ...752...32W,2019ApJ...881L...1F}, so rotational angular momentum is often considered insufficient to generate an outflow if a RSG collapses following a FSN. It is often assumed that this implies that the envelope can accrete spherically onto the newborn BH.  Indeed, the optically-quiet disappearance of RSGs in FSNe has motivated the, now, decade-long study of \citet{2008ApJ...684.1336K} monitoring $10^6$ RSGs in nearby galaxies and waiting for one or two to disappear.  Their strongest candidate for a FSN is N6946-BH1 in which a $\sim 25\msun$ RSG brightened for several months before dimming to an optical luminosity that is far below that of the progenitor \citep{2017MNRAS.468.4968A,2021MNRAS.508..516N}.  The event is qualitatively similar to the results of \citet{2013ApJ...769..109L} for faint transients produced by neutrino-induced mass-loss in RSG collapse but further work is needed to understand the details of the N6946-BH1 light curve.

It turns out that non-zero total angular momentum is not necessary to prevent spherical (quiescent) infall of RSGs and yellow supergiants (YSGs). The random velocity field in stellar convective zones carries finite angular momentum. This idea was first suggested by \citet{2014MNRAS.439.4011G} and studied numerically by \citet{2016ApJ...827...40G} in the context of ``jittering-jet"-driven SN \citep[e.g., ][]{2010MNRAS.401.2793S,2011MNRAS.416.1697P}. Indeed, \citet{Quataert2019} and \citet{2022MNRAS.511..176A} have shown that the convective hydrogen envelopes of even non-rotating RSGs and YSGs carry sufficient angular momentum to prevent spherical accretion, and may be capable of generating a luminous transient \citep[see also][]{2022ApJ...929..156G}. Hydrodynamical simulations have shown that the mean angular momentum associated with convection in the zones interior to the hydrogen envelope (e.g. the helium layer) is at least two orders-of-magnitude smaller than the angular momentum associated with the inner-most stable circular orbit (ISCO) of the BH formed from the collapsing core \citep{2016ApJ...827...40G,Quataert2019,2022MNRAS.511..176A}. Thus, we assume those layers can accrete quiescently onto the BH and instead focus on infall of the convective hydrogen envelope in our study here.

In \citet[][hereafter, Paper I]{2022MNRAS.511..176A}, we modeled the power-law, convective hydrogen envelope of a non-rotating supergiant, then followed the inflow of the convective material during the initial collapse of the star.  There, we showed that the random velocity field due to convection gives rise to a mean specific angular momentum at each radius, with magnitude $\jrand(r)$, that is a factor of $\sim$7 - 50 times larger than $\jisco$, the specific angular momentum associated with the innermost stable circular orbit (ISCO) of the BH formed inside the star.   We also showed that, during the initial collapse, $\jrand(r)$ in each shell is largely conserved such that the circularization radius, $\rcirc$, of the material in each shell prior to collapse is indeed the radius at which centrifugal deflection should begin to occur as the gas falls in toward the BH.  In Paper I, we simulated a factor of $\sim$ 20 in radius of the hydrogen envelope out to the photosphere of the star, so we were only able to follow the infall of the gas down to a radius $\sim 60$ times larger than $\rcirc$ of the material.   In the present study, we follow the infall of the gas to yet smaller radii. Critically, we resolve the flows at $\rcirc$ so that we can understand the outflows that are generated by the infall of the convective envelope and the extent to which the collapse of a non-rotating supergiant with a convective hydrogen envelope may generate a luminous transient.

This paper is organized as follows. Our simulation methods are described in Section \ref{sec:methods}.  Section \ref{sec:results} presents the results from our suite of simulations. Section~\ref{sec:extrap} extrapolates our simulation results in time to apply to collapse of real RSGs. In Section~\ref{sec:obs}, we estimate the observational implications of our results. We summarize our results in Section~\ref{sec:conclusion}.

\section{Numerical methods}
\label{sec:methods}
Paper I modeled the collapse of the convective envelope of a non-rotating supergiant in 3D, simulating a factor of $\sim 20$ in radius of the outer hydrogen envelope. The collapse of the envelope was facilitated by the introduction of an absorbing sink at the center of the star, which removed hydrostatic pressure support and permitted accretion at small radii without feedback.  In those simulations, the radius of the absorbing sink, $\rsink$, was a factor of $\sim 60$ larger than the circularization radius of the accreting material, $\rcirc = j^2/GM$.  Here we perform a set of ``zoom-in'' simulations in which we use a collapse simulation similar to those of Paper I to provide the initial conditions but we follow the inner accretion flow to much smaller radii.  Critically, we have $\rsink < \rcirc$ and we are able to resolve the flow at scales where rotation becomes dynamically important.  In the remainder of this section, we describe our simulation method including our procedure for initializing the grid using data from the ``parent'' simulation of Paper I. 

We use the Eulerian hydrodynamics code \ath\footnote{Version 19.0, \href{https://princetonuniversity.github.io/athena}{https://princetonuniversity.github.io/athena}}
\citep{2020ApJS..249....4S} to solve the equations of inviscid hydrodynamics, 
\beq
\frac{\upartial \rho}{\upartial t} + \nabla\cdot(\rho \bmath{v}) = 0
\eeq
\beq
\frac{\upartial (\rho{\bmath v})}{\upartial t} + \nabla\cdot(\rho \bmath{v}\bmath{v} + P \mathbfss{I}) = - \rho \nabla \Phi 
\label{eq:momentum}
\eeq
\beq
\frac{\upartial E}{\upartial t} + \nabla\cdot\big[(E + P)\bmath{v}\big] = -\rho \bmath{v}\cdot \nabla \Phi,
\label{eq:energy}
\eeq
employing third-order time integration (Runge-Kutta, \texttt{integrator = rk3}), third-order spatial reconstruction (piecewise parabolic, \texttt{xorder = 3}), and the Harten-Lax-van Leer contact Riemann solver (\texttt{--flux hllc}). In the above equations, $\rho$ is the mass density, $\rho \bmath{v}$ is the momentum density, $E = \epsilon + \rho \bmath{v}\cdot\bmath{v}/2$ is the total energy density, $\epsilon$ is the thermal energy density, $P$ is the gas pressure, \mathbfss{I} is the identity tensor, and $\Phi = -GM/r$.  As shown in Paper I, we expect all of the material interior to the hydrogen envelope to have $\jrand \ll \jisco$ and to accrete spherically onto the core.  All of this mass is contained in the point mass, $M$. We do not evolve $M$ in time because the total (hydrogen) gas mass accreted during the simulation is negligible compared to $M$.

We adopt an ideal gas EOS, $\epsilon = {P}/{(\gamma -1)}$, with adiabatic index $\gamma = 1.4762$.  In Paper I, the convective hydrogen envelope was modeled as a polytrope with a density profile $\rho \propto r^{-b}$  in hydrostatic equilibrium with a Plummer potential. The density slope $b = 2.1$ matches the hydrogen envelope of a 16.5$\msun$ RSG modeled with \mesa \citep{2011ApJS..192....3P,2013ApJS..208....4P,2015ApJS..220...15P,2018ApJS..234...34P,2019ApJS..243...10P} but values of $1.5 \lesssim b \lesssim 2.5$ are typical for RSGs and YSGs.  The $T \propto r^{-1}$ temperature profile of our ideal-gas, power-law envelope was smoothly dropped to an isothermal atmosphere beyond the photosphere of the star. The value $\gamma  = 1.4762$ gives a flat entropy profile in the polytropic envelope of the star. In Paper I, we generated convection by heating at small radii and cooling at the photosphere and we allowed sufficient time for the model to reach thermal equilibrium.  We achieved convective Mach number profiles typical of RSGs and YSGs allowing our dimensionless setup to be scaled to a wide range of supergiants (see figs. 1 and 6 of Paper I for comparisons between our \ath model and the \mesa model).  The thermal-equilibrium, convective envelope provided the initial conditions for the collapse simulations in that work as well as for the zoom-in collapse simulations we study here.  For numerical reasons, as discussed below, we start the collapse in the ``parent'' setup of Paper I, then use snapshots from that simulation to initialize the zoom-in models we study here. 

Our convective star is non-rotating and the angular momentum field associated with the random convective velocity field is randomly oriented, so there is no natural symmetry axis that would otherwise motivate the use of a spherical-polar grid.  Our simulations are thus performed in a Cartesian grid.  Other details about the computational domain are given later in this section.   

Our implementation does not have an explicit cooling function. This is a valid assumption as our inflow is advection-dominated and radiately-inefficient with  accretion rates of $\sim$0.1 - 10 $\msun$ / yr.  The material is optically thick so radiation is trapped yet the accretion rates are too low for neutrino cooling.

We note that the present paper is restricted to purely hydrodynamical simulations. Future simulations will include magnetic fields which are undoubtedly important for angular momentum redistribution.

\subsection{Boundary conditions}
At the outer boundaries of the simulation domain, we employ \athns's zero-gradient \texttt{outflow} boundary condition. We stop our simulations before any material from the outer boundaries can interact with the physically important part of the domain.  

The center of the star is located at the origin of our Cartesian grid. There, we place an absorbing sink of radius $
\rsink$.  Inside the sink, the fluid velocities are set to zero and the pressure and density are floored to values of $10^{-5}$ (or smaller) times the typical values expected from extrapolating the pressure and density profiles from the parent simulations (see Sec.~\ref{sec:initial_zoom_in_grid}). The low pressure and density allow material to flow unimpeded into the sink, which represents accretion onto the BH without feedback.

\subsection{Code units}
\label{sec:units}
Our code units adopt $GM = 1$ and length unit $r_0 = 1$, which yields velocity units of
\beq 
v_0 \equiv ({GM/r_0})^{1/2} = 1
\eeq
 and time units of
 \beq
 t_0 \equiv (r_0^3 / GM)^{1/2} = 1.
 \eeq
The Keplerian specific angular momentum at $r_0$ is $j_0 \equiv \sqrt{GM r_0} = 1$.

In code units, the density profile in the polytropic envelope of the pre-collapse star is $\rho(r) = \rho_0 (r/r_0)^{-b}$, with density unit $\rho_0$ and $b=2.1$. The photosphere of the model is located at $r = 6.0r_0$. Given our $\gamma$-law equation of state, this yields a characteristic adiabatic sound speed  of $c_{\rm s,0} \equiv v_0\sqrt{\gamma/(b+1)}$. Units of gas mass are $m_0 = \rho_0 r_0^3$ and mass accretion rates are in units of $\dot{m}_0 \equiv m_0/t_0 = {\rho_0 r_0^3}/{t_0} = \rho_0 r_0^{3/2}(GM)^{1/2}$.  Units of energy are $\mathcal{E}_0 \equiv \rho_0 r_0^3 v_0^2 = GM \rho_0 r_0^2$.

\subsection{Initial conditions}
\label{sec:initialization}

\subsubsection{Parent simulation}
\label{sec:parentsim}
We initialize the gas variables in the grid using a snapshot from a collapse simulation similar to those of Paper I.  The simulation we use is identical to Run 2 from Paper I except that it employs a sink radius of $0.005$ $r_0$ (instead of 0.08 $r_0$) and 8 levels of static mesh refinement on top of the base resolution (instead of 5 levels).  Using a somewhat smaller sink in this ``parent'' simulation allows us to save some compute time in the zoom-in simulations but does not change the results in either the parent or the zoom-in simulations.  

\begin{figure}
\centering
	\includegraphics[width=0.95\columnwidth]{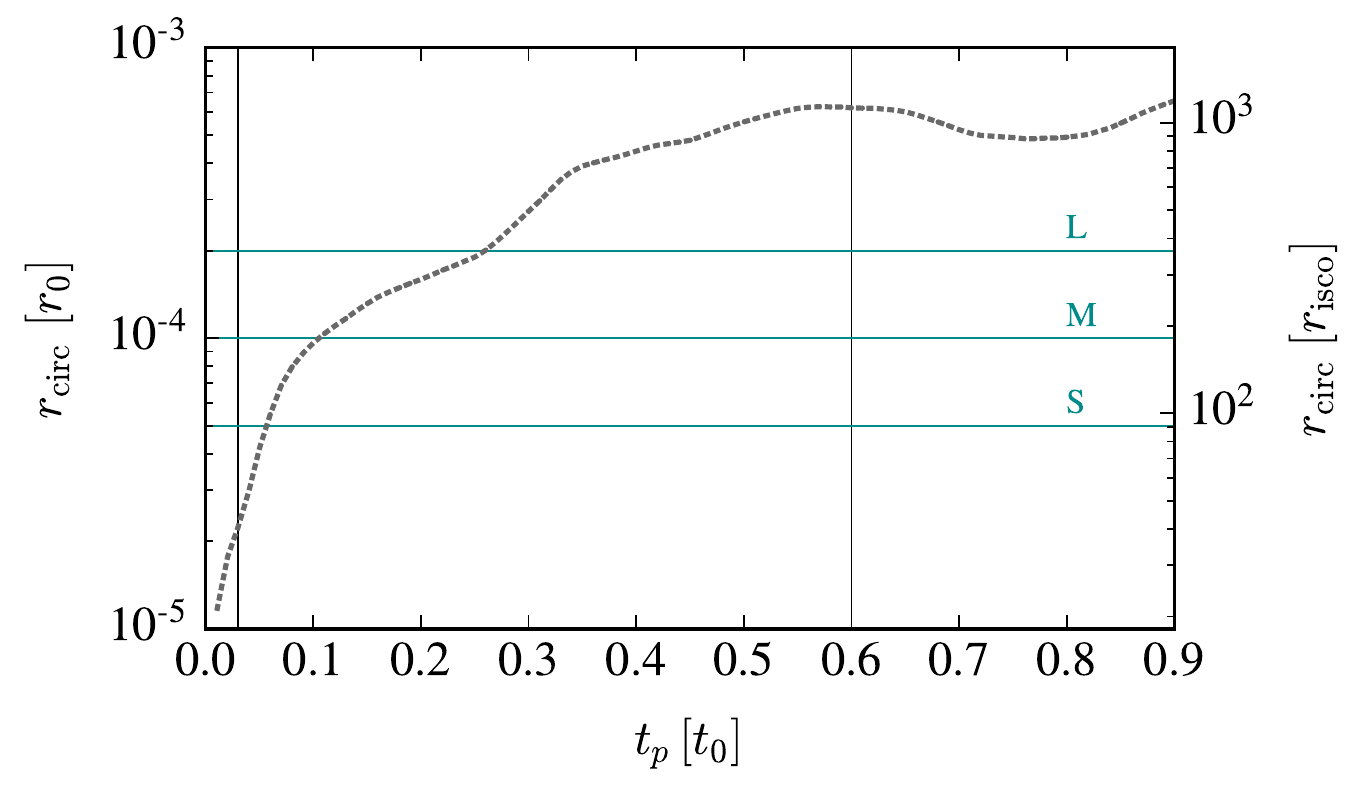}
    \caption{Circularization radius of the accreted material as a function of time in the parent simulation (similar to Run 2 of Paper I) that is used to initialize the zoom-in simulations studied in this paper. The value of $r_{\rm circ}$ shown here is the circularization radius of the material that accretes into a sink with radius $r_s = 0.005 r_0$ in the parent simulation.  Since $\rsink > r_{\rm circ}$, there is no deflection of material due to angular momentum in the parent simulation.  The purpose of the zoom-in simulations presented here is to employ $\rsink < \rcirc$ in order to follow the flow down to scales where angular momentum becomes important. The three horizontal lines mark the values of $\rsink$ we adopt across our suite of zoom-in simulations. For reference, the right-hand $y$-axis gives $\rcirc$ in units appropriate for our 16.5$\msun$ RSG model, relative to the ISCO of the 6$\msun$ BH.}
\label{fig:rcirc}
\end{figure}

Fig.~\ref{fig:rcirc} shows the circularization radius of the accreted material, $r_{\rm circ}$, as a function of time since the start of collapse in the parent model, $t_p$. The vertical lines in the figure show the two different snapshots we use to initialize our suite of zoom-in simulations.  Our primary simulations are initialized using the snapshot at $t_p = 0.6 t_0$, when $r_{\rm circ} = 6.25$ $\times 10^{-4} r_0$. We choose this time because it represents the time at which the collapsing gas in the parent simulation reaches its saturated specific angular momentum and circularization radius (Fig.~\ref{fig:rcirc}). There is no physical reason preventing us from beginning from a snapshot at an earlier time with smaller $r_{\rm circ}$. However, it would increase the computational cost of each zoom-in simulation as it would require both a smaller sink (shorter time-step) and higher refinement (shorter time-step and more compute zones) in order to resolve the flow at $r_{\rm circ}$.  To show that using a snapshot from a later time does not significantly change our results, we also run two half-resolution simulations using $t_p = 0.03t_0$ (the first vertical line in the figure).

\subsubsection{Initialization of the zoom-in grid}
\label{sec:initial_zoom_in_grid}
The parent simulation snapshot that we use to initialize the zoom-in grid contains the density, pressure, and Cartesian velocity components as a function of cell-centered coordinates $x$, $y$, and $z$ for the full parent grid, which extends to $\pm 20$ $r_0$ in each Cartesian direction. The domain in our zoom-in models is smaller, extending either to $\pm 2.5$ or $\pm 7.5$ $r_0$ in each direction, depending on the simulation (see Sec.~\ref{sec:refinement_and_parameters}). Interpolating the gas variables is not necessary because the lowest levels of refinement in the zoom-in grids have an identical grid structure to the highest levels of refinement in the parent grid. The cell-centers align to numerical precision, so nearest-neighbor is sufficient.  On top of those lowest levels of refinement, additional nested refinement levels are added in the zoom-in simulations (Sec.~\ref{sec:refinement_and_parameters}). 

The parent simulation employed a sink radius of $0.005r_0$. Within the sink region, the density and pressure were floored and the velocities were set to zero.  After filling the zoom-in grid from the parent data, we overwrite the gas variables in the grid zones with $r \le 0.008r_0$ by extrapolating the density, pressure, and radial velocity from $0.008r_0$ (slightly larger than the original sink region) down to the new sink radius of the zoom-in sim ($\rsink \le 0.0002 r_0$).  The extrapolated values are zero-angular-momentum Bondi solutions with the normalization set so that spherically-averaged profiles of $\rho$, $P$, and the radial velocity, $v_r$, are continuous and smooth at $r = 0.008r_0$. Inside the zoom-in sink, the velocities are set to zero and we floor the pressure and density to $\sim 10^{-6}$ and $10^{-5}$, respectively, times the extrapolated values of those quantities at $\rsink$.  Filling the original vacuum region with the free-fall solutions helps with numerical stability at $\sim$ the old sink radius during the first few time steps. It also prevents a rarefaction wave from propagating out from the new sink radius and interacting with the physical material initialized at $r > 0.008r_0$.  

\begin{figure}
\centering
	\includegraphics[width=0.95\columnwidth]{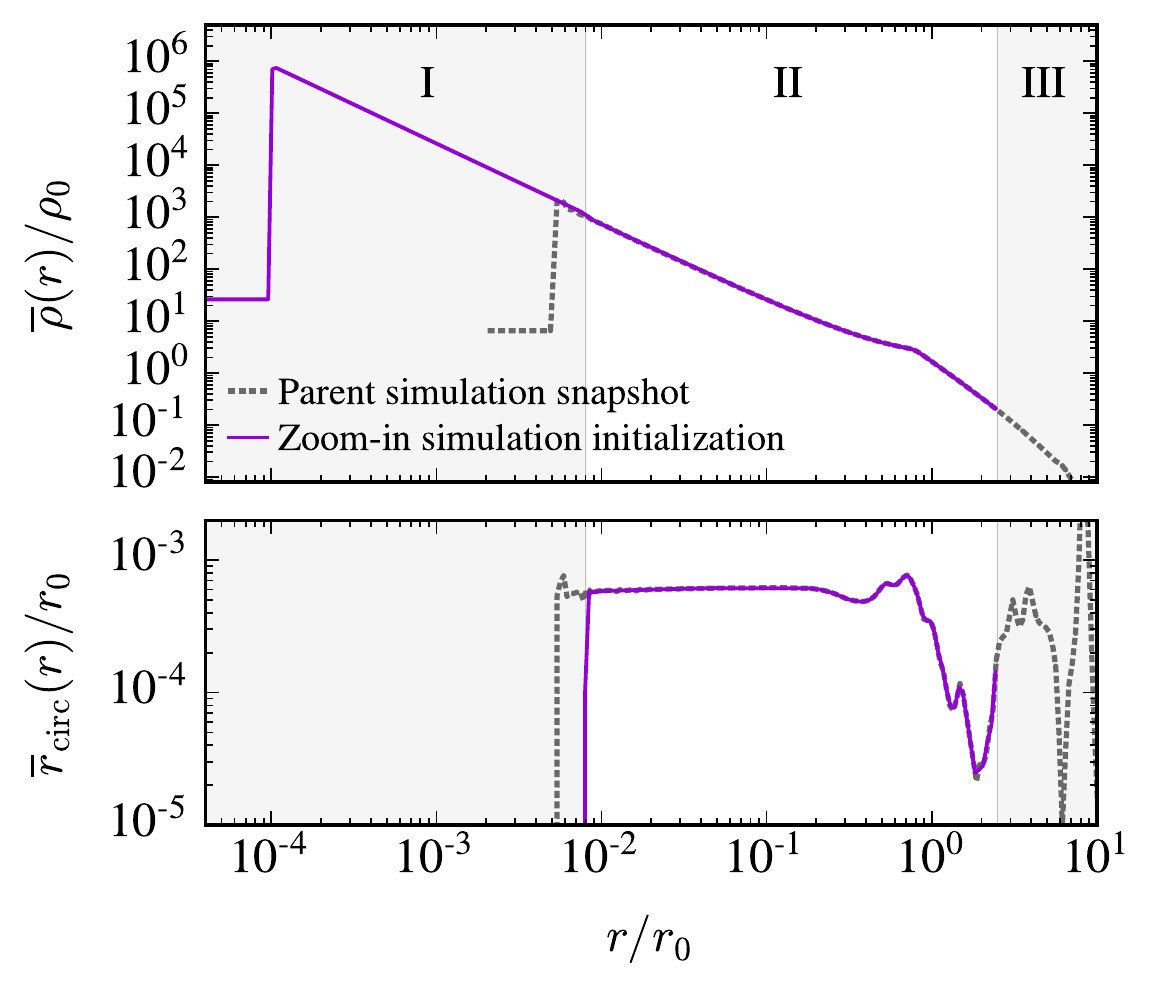}
    \caption{Spherically-averaged profiles of mass density (top) and circularization radius (bottom) from the 3D grids of the parent simulation at $t_p = 0.6t_0$ (dotted lines) and of the zoom-in grids at initialization (solid, purple lines).  The zoom-in model shown is Model M, but models S, M, L and \Mhalf would look identical. Region I: The parent simulation data is overwritten with zero-angular-momentum Bondi extrapolations of the density, pressure and radial velocity out to a slightly larger radius than the sink region in the parent simulation. Region III: The part of the parent domain that falls outside of the zoom-in domain. Region II: The zoom-in simulation is initialized using the parent simulation solution. This is the physically important material in the grid, whose infall we follow to small radii in the zoom-in simulations.}
\label{fig:initial_profile}
\end{figure}

The grey, dashed lines of Fig.~\ref{fig:initial_profile} show spherically-averaged profiles of density ($\overline{\rho}$, upper panel) and circularization radius ($\overline{r}_{\rm circ} = \overline{\jmath}^2/GM$, lower panel) from the $t_p = 0.6t_0$ snapshot of the parent simulation. The step in both variables occurs at the original sink radius of $0.005r_0$.  The purple, solid lines show profiles of our fiducial zoom-in simulation (Model M; see Sec.~\ref{sec:refinement_and_parameters}) at initialization.  Region III is the portion of the parent simulation that falls outside of the zoom-in domain. Region II is the physical part of the zoom-in grid initialized from the parent snapshot.  Region I is the Bondi extrapolation region. The density step in the purple line in Region I occurs at the new sink radius (in this model, $\rsink = 10^{-4}r_0$). The material in Region I carries no angular momentum ($\rcirc < 10^{-50} r_0$). It accretes spherically and supersonically by $t < 10^{-3} t_0$ without influencing the physically important material flowing in from Region II.  

\begin{figure}
\centering
	\includegraphics[width=1.0\columnwidth]{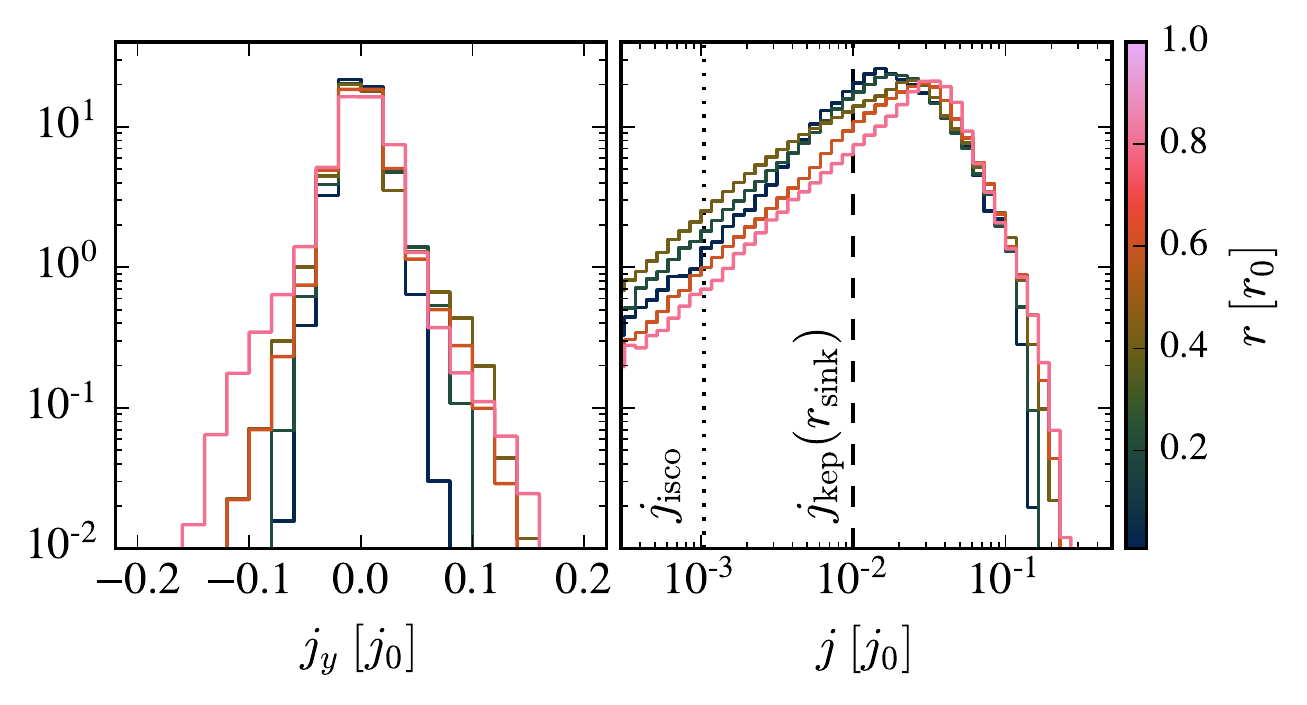}
    \caption{Histograms of the $y$-component (left) and the magnitude (right) of specific angular momentum in radial shells between $0.008$ and $1.0r_0$ in our 3D grid at initialization of all $t_p = 0.6t_0$ models. (distributions of the $x$- and $z$-components are similar to $y$). Left panel: At initialization, the gas has mean specific angular momentum with magnitude $\approx 0.024 j_0$ between $0.008$ and $1.0r_0$ (square root of the purple line in the lower panel of Fig.~\ref{fig:initial_profile}), but the widths of the $j_x$, $j_y$, and $j_z$ distributions at all radii are almost an order-of-magnitude larger than the mean. Right panel: The material has a wide range of $j$ (and, thus, circularization radii). At every radius, only a small fraction of the material has $\rcirc < \rsink$ (material to the left of the dashed line). For reference, we include the dotted line at $j_{\rm isco} = 1.05 \times 10^{-3}j_0$ which has been scaled for our \mesa RSG model.}
\label{fig:hist}
\end{figure}
We emphasize that the angular momentum field arising from the random velocities in the convective envelope is very different from that of the rotationally-supported flows usually considered in massive stellar collapse or in radiatively inefficient BH accretion flows.  Random, convective flows have a wide distribution of specific angular momentum vectors, $\vec{\jmath}$. The individual parcels in each shell sample a wide distribution of circularization radii. Thus, the parcels in each shell should be deflected from spherical infall at a wide range of radii.  Critically, the {\it direction} of $\vec{\jmath}$ also varies within each shell, so there is no fixed rotation axis like the flows studied in, e.g., collapse of rotating stars.  The left panel of Fig.~\ref{fig:hist} shows distributions of the $y$-component of $\vec{\jmath}$ at initialization of the $t_p = 0.6 r_0$ models. Each curve corresponds to a different radial shell between $0.008$ and $1.0$ $r_0$.  The distributions of $j_x$ and $j_y$ are very similar, so are not shown.  The right panel of the figure shows distributions of the magnitude of $\vec{\jmath}$ for the same radial shells.  The mean value of $j \approx 0.024 j_0$ between $0.008$ and $1.0r_0$ (see Fig.~\ref{fig:initial_profile}), but the $j_y$ distributions are almost a factor of 10 wider at all radii. As shown in the right panel, most of the material has $j$ larger than the Keplerian specific angular momentum at the sink radius (the vertical, dashed line), and will be deflected from spherical infall outside of $\rsink$.  
Roughly 6\% of the gas between 0.008 and 1.0 $r_0$ has $j <j_{\rm Kep}(\rsink)$ and, assuming $j$ is fixed during infall, should be able to accrete into the sink.  We will see later that in fact most of this low angular momentum material does not accrete as it is advected with the bulk of the mass, which has $j > j_{\rm Kep}(\rsink)$. 

\subsection{Summary of simulations performed}
\label{sec:refinement_and_parameters}

\begin{table*}
	\centering
	\caption{Table of simulations performed.  Model M is our fiducial model. See Table \ref{tab:units} for conversions to physical units.}
	\begin{tabular}{lcccccccc} 
	  \hline
	  Model     & $r_{\rm sink}$$^{a}$   & Domain$^{b}$  & Base$^{c}$ & Refine$^{d}$ & $\Delta x_{\rm min}$$^{e}$ &  &$t_p$ $^{f}$ &  $t_{\rm max}$$^{g}$ \\
	  Name & $[r_0]$            & Volume  $[r_0^3]$   & Cells & Levels        & $[r_0]$     &$r_{\rm sink}/\Delta x_{\rm min}$ &$[t_0]$ & $ [t_0]$ \\
	  \hline
	  \hline
	  {\it Full-resolution models} \\
	  \hline
	  S      & $5\times 10^{-5}$ & 5$^3$ &128$^3$     & 13  & $4.8\times 10^{-6}$ &10.49 &0.6 & 0.4 \\
	  M      & $1\times 10^{-4}$ & 5$^3$ &128$^3$     & 12  & $9.5\times 10^{-6}$ &10.49 &0.6 & 0.4  \\
	  L      & $2\times 10^{-4}$ & 5$^3$ &128$^3$     & 11  & $1.9\times 10^{-5}$ &10.49 &0.6 & 0.4 \\
	  \hline
	  {\it Half-resolution models} \\
	  \hline

	  \Mhalf   & $1\times 10^{-4}$ & 5$^3$  &64$^3$  & 12  & $1.9\times 10^{-5}$ & 5.24 &0.6  & 0.4 \\  
	  \Mhalftp & $1\times 10^{-4}$ &15$^3$  &96$^3$  & 13  & $1.9\times 10^{-5}$ & 5.24 &0.03 & 2.5 \\
	  \Shalftp & $5\times 10^{-5}$ &15$^3$  &96$^3$  & 14  & $9.5\times 10^{-6}$ & 5.24 &0.03 & 0.46 \\

	  \hline
	\end{tabular}
	\vspace{1ex}
	{\par \raggedright {\it Notes.} \\
	$^{a}$ Radius of low-pressure sink activated at the start of the simulation. The model names reference $r_{\rm sink}$ (S=small, M=medium, L=large). \\ 
	$^{b}$ Total volume of the simulation domain, which is centered at $x=y=z=0$.\\
	$^{c}$ Number of grid cells in the base, unrefined grid.\\ 
	$^{d}$ Number of static mesh refinement levels on top of the base resolution. When we change $r_{\rm sink}$ by a factor of 2, we add or subtract one refinement level in the innermost region to fix the number of grid cells across the sink radius, $r_{\rm sink}/\Delta x_{\rm min}$. \\ 
	$^{e}$ Width of each grid cell in the highest refinement region. \\ 
	$^{f}$ Snapshot in time in the parent simulation used to initialize the zoom-in simulation. Times in the zoom-in simulation, $t$, all start at zero independent of $t_p$.\\ 
	$^{g}$ Simulation run time in the zoom-in simulation (starting at $t = 0$). \\ \par}
\label{tab:models}
\end{table*}

Our suite of zoom-in simulations is listed in Table~\ref{tab:models}.  These models vary the radius of the absorbing sink, $\rsink$, the base grid resolution, and the parent snapshot used to initialize the grid (parameterized by $t_p$, the time since collapse in the parent simulation).   Different simulations may also employ a different number of refinement levels or a different domain size, depending on $\rsink$ or $t_p$, as described below.

The sink radii for our models are chosen to ensure that $\rsink$ is sufficiently smaller than the mean circularization radius of the infalling material in order to study the influence of the convective angular momentum on the accretion flow.  Ideally, our simulations would have $\rsink \sim r_{\rm horizon}$ in order to capture all of the available accretion energy associated with the rotating flows we are simulating. This is, however,  impractical: the time-step for such a simulation would be $\sim (r_{\rm horizon}/10)/c \sim 10^{-5}$ sec while we are interested in running for a fraction of the free-fall time of the red-giant $\sim$ months $\sim 10^{11}$ time-steps in order to assess convergence (in time) of the total energy supplied to outflows and the total mass ejected during the collapse of the RSG envelope. Our simulations are thus severely time-step limited. Our strategy is to choose values of $\rsink$ that are small enough to capture the circularization of the majority of the gas while at the same time large enough that we can run our simulations for nearly a free-fall time of the progenitor star. We then check how our results change with modest variations in $\rsink$. A second motivation for this choice is that simulations with yet smaller $\rsink$ should ideally include magnetic fields to capture angular momentum redistribution in the infalling gas. We defer such calculations to future work.

Our fiducial simulation is Model M, which adopts $\rsink = 1 \times 10^{-4}$ $r_0$ and is initialized using the parent snapshot with $t_p = 0.6$ $t_0$.  The sink radius is halved in Model S ($\rsink = 5\times 10^{-5}$ $r_0$) and is doubled in Model L ($\rsink = 2\times 10^{-4}$ $r_0$). The simulation domain in these models extends to $\pm 2.5$ $r_0$ in $x$, $y$, and $z$ (for a total simulation volume of $125 r_0^3$), allowing us to safely run all models for $0.4t_0$ (after this time, the outflow in Model S approaches the outer boundary).  The horizontal lines in Fig.~\ref{fig:rcirc} show the three values of $\rsink$ we adopt in these models.   At $t_p = 0.6 t_0$ (the second vertical line in Fig.~\ref{fig:rcirc}), $\rcirc > \rsink$. Our models with $t_p=0.03t_0$ (described below) adopt sink sizes S and M. For these models, $\rsink > \rcirc$ until $\sim 0.1 t_0$ after initializing the zoom-in simulation.

Model M applies 12 levels of static mesh refinement on top of the base grid of $128^3$ zones to increase resolution towards the origin (the center of the star).  The refinement transitions occur where $r = (x^2 + y^2 + z^2)^{1/2} =$ $\{$1.25, 0.625, 0.3125, 0.156, 0.078, 0.039, 0.0195, 0.0098, 0.0049, 0.0024, 0.0012, 0.0006$\}$ $r_0$.    The zone size is reduced by a factor of two at each jump in refinement, giving a cell volume of $(\Delta x_{\rm min})^3 = (9.5 \times 10^{-6} r_0)^3$ where $r \le 0.0006$ $r_{0}$. This refinement results in 10.5 zones across the sink radius.   The grid is the same in Models S and L except that a refinement region is added or subtracted to maintain 10.5 zones across $\rsink$.  The 13th refinement level is added to Model S in the  region with $r \le 0.0003$ $r_0$.  Model L omits the 12th refinement region of Model M so that the highest refinement region in Model L is $r \le 0.0012$ $r_{0}$.  We run one additional model with $t_p = 0.6$ $t_0$, Model \Mhalfns, which isolates the effect of lowering the resolution everywhere by a factor of two.  The model is identical to Model M except that it has a base resolution of $64^3$ zones.

To test our choice of initializing the above models from $t_p = 0.6t_0$, we run Model \Mhalftp at half our fiducial resolution and with $t_p = 0.03t_0$.  This start time is essentially $t_p = 0$, but it allows for the small ramp time of $0.03t_0$ in the parent simulation during which the pressure and density inside the parent sink are lowered smoothly to their target values (in order to avoid hydro errors in the first few time steps).  The ramp time delay has no effect on any of our results. Model M was run for 0.4$t_0$ once initialized at 0.6$t_0$ after the start of the collapse in the parent simulation.  To reach this same point, Model \Mhalftp has to be run for at least 1.0 $t_0$.  Because the outflow is starting earlier, this model requires a larger domain to avoid interaction between the outflow and the outer boundary.  The domain extends to $\pm 7.5 r_0$ in each direction.  The base resolution and location of the refinement transitions are chosen so that the grid structure is identical to Model \Mhalf inside of $\pm 2.5$ $r_0$.  Outside of that region, the grid is similar to that of the parent simulation.   

One final model checks that the behavior with changing $\rsink$ that we find at high resolution holds at half resolution and with $t_p = 0.03t_0$.  Model \Shalftp is identical to Model \Mhalftp except that $\rsink = 5\times 10^{-5}$ $r_0$ and one additional level of refinement is added (same as in Model S). 

\subsection{Scaling code units to a stellar model}
\label{sec:scaledunits}
Our dimensionless setup can be scaled to a particular RSG or YSG envelope by associating the photosphere of the model with the photosphere radius in our setup, $r \approx 6r_0$.  From the stellar model, identify the photosphere radius, $r_{\rm ph}$, the mass interior to the hydrogen envelope, $M_\bullet$, and the density at $r_0 = r_{\rm ph} / 6$, $\rho_\star$.  Setting $M = M_\bullet$, $r_0 = r_{\rm ph} / 6$, and $\rho_0 = \rho_\star$ yields the following relations to scale code units to the stellar model:
\beq
r_0 = 1.16\times10^{13} \bigg(\frac{r_{\rm ph}}{1000\rsun}\bigg) {\rm cm},
\label{eq:r0_star}
\eeq

\beq
t_0 = 0.034 \bigg(\frac{r_{\rm ph}}{1000\rsun}\bigg)^{{3}/{2}}\bigg(\frac{10 \msun}{M_\bullet}\bigg)^{{1}/{2}} \text{yr},
\label{eq:t0_star}
\eeq

\begin{align}
j_0  =  1.24 \times 10^{20} \bigg(\frac{r_{\rm ph}}{1000\rsun}\bigg)^{1/2}\bigg(\frac{M_\bullet}{10\msun}\bigg)^{1/2} 
\frac{{\rm cm}^2}{{\rm s}},
\label{eq:j0_star}
\end{align}

\beq
\dot{m}_0 = 2.27 \bigg(\frac{\rho_\star}{10^{-7} \frac{\text{g}}{\text{cm}^{3}}}\bigg) \bigg(\frac{r_{\rm ph}}{1000\rsun}\bigg)^{{3}/{2}} \bigg(\frac{M_\bullet}{10 \msun}\bigg)^{{1}/{2}}\frac{\msun}{{\rm yr}},
\label{eq:mdot0_star}
\eeq
and
\beq
\mathcal{E}_0 = 1.795 \times 10^{46} \bigg(\frac{\rho_\star}{10^{-7} \frac{\text{g}}{\text{cm}^{3}}}\bigg) \bigg(\frac{M_\bullet}{10 \msun}\bigg)\bigg(\frac{r_{\rm ph}}{1000\rsun}\bigg)^{2} {\rm erg}.
\eeq
As in Paper I, we adopt $r_{\rm ph} = 840 \rsun$, $M_\bullet = 6 \msun$, and $\rho_\star = 4.0 \times 10^{-7}$ g cm$^{-3}$ when scaling code units to a physical stellar model.  These are from a 16.5$\msun$ RSG model, computed with \mesans, which was evolved from an 18$\msun$ star until near the end of oxygen burning (profiles of the \mesa model can be found in figs. 1 and 6 of Paper I). Table~\ref{tab:units} gives the simulation parameters of Model M in these scaled units.
 
A useful quantity to have in our code units is the radius of the ISCO for a non-spinning BH with mass $M_\bullet$,
\beq
\frac{r_{\rm isco}}{r_0} = 7.7\times10^{-7} \bigg(\frac{M_\bullet}{10 \msun}\bigg)\bigg(\frac{r_{\rm ph}}{1000\rsun}\bigg)^{-1}.
\eeq
For our 16.5$\msun$ \mesa model, $r_{\rm isco} = 5.5 \times 10^{-7} r_0$.

\begin{table}
	\centering
	\caption{Scaled units and simulation parameters for a model RSG.}
	\begin{tabular}{ll} 
	  \hline
	  \multicolumn{2}{l}{Scaled Units}\\
	  \hline
	  $r_0$ & 140 $\rsun$\\
	  $t_0$ & 12.4 days \\
	  $j_0$ & 8.8 $\times$ 10$^{19}$ cm$^2$/s\\
	  $\dot{m}_0$ & 5.43 $\msun$/yr \\
	  $\mathcal{E}_0$ &3 $\times$ 10$^{46}$ erg  \\
	  \hline
	  \multicolumn{2}{l}{Scaled Model M Parameters}\\
	  \hline
	  $\rsink$ & 0.014 $\rsun$\\
	  Domain Volume & (700 $\rsun$)$^3$ \\
	  $\Delta x_{\rm min}$ & 0.0013$\rsun$ \\
	  $t_p$ & 7.5 days\\
	  $t_{\rm max}$ & 5.0 days\\
	  \hline
	\end{tabular}
	\vspace{1ex}
	{\par \raggedright {\it Notes.}  Scaled quantities adopt $r_{\rm ph} = 840 \rsun$, $M_\bullet = 6 \msun$, and $\rho_\star = 4.0 \times 10^{-7}$ g cm$^{-3}$ obtained from a 16.5$\msun$ RSG modeled with \mesans. The mass of the hydrogen envelope for this model is $M_{\rm env} = 10.5 \msun$.  Simulation parameters are for Model M (described in Sec.~\ref{sec:refinement_and_parameters}). The simulation begins at time $t_p$ after the collapse begins in the parent simulation and runs for a time $t_{\rm max}$. \\ \par}
\label{tab:units}
\end{table}

\section{Simulation Results}
\label{sec:results}

In this section, we describe the results of the simulations listed in Table \ref{tab:models}.  Section~\ref{sec:flow} presents qualitative features of the gas flows for our fiducial simulation, Model M.  Secs.~\ref{sec:accretion} and \ref{sec:outflow} present quantitative results from all simulations with $t_p = 0.6 r_0$, namely, Models S, M, L, and \Mhalfns.  Lastly, Sec.~\ref{sec:tparent} presents the results of Models \Mhalftp and \Shalftpns, which show that initializing the zoom-in simulation from an earlier time in the parent simulation yields similar results.
 
\subsection{Overall structure}
\label{sec:flow}
As discussed in Sec.~\ref{sec:initial_zoom_in_grid} and shown in the histograms of Fig.~\ref{fig:hist}, random convective flows have a broad distribution of angular momentum vectors.  This is very different from the collapse of rotating stars for which the rotation axis is fixed in time. In that case, simulations show that the outflows are collimated along a somewhat fixed axis and accretion happens in a well-defined plane perpendicular to the rotation axis. In our simulations, we find that outflows are indeed generated because of the reservoir of angular momentum in the convective envelope but the flows are different from the rotating-star case.  Namely, we see no evidence of disk accretion nor a defined outflow direction. Instead, accreting gas approaches the sink from randomly-oriented directions over time.  We highlight these qualitative features of our simulations in the next few figures. 

\begin{figure*}
\centering
	\includegraphics[width=0.95\textwidth]{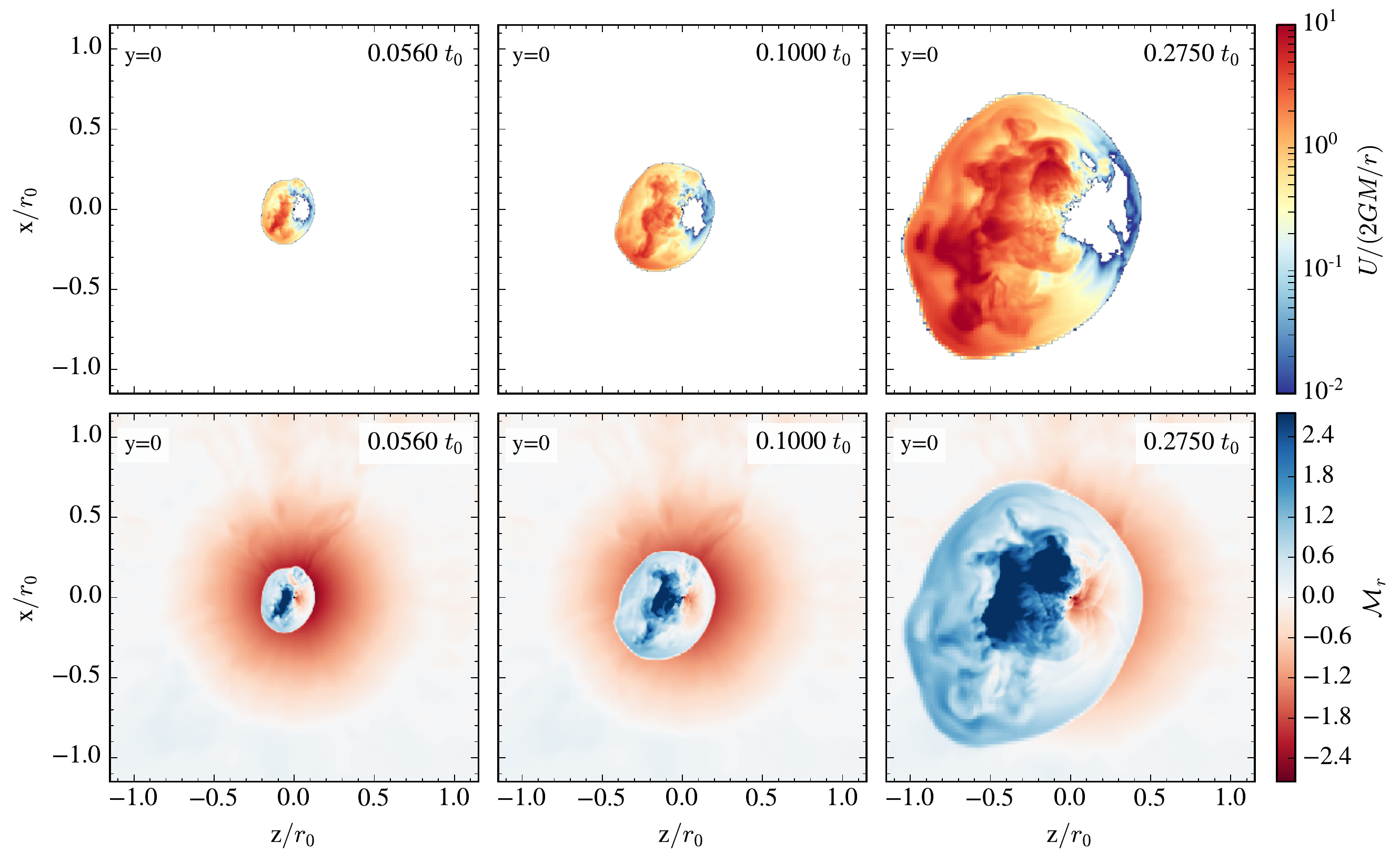}
    \caption{Slices through the $y=0$ plane over time of specific energy (top row) and local radial Mach number (bottom row) for Model M.  In the top row, material with $U < 0$ is shown in white. In the bottom row, the convective background appears mostly white as the convective Mach numbers are only $\sim 0.1-0.3$. The red background material in the bottom row is the gas that was already falling in at initialization ($t_p = 0.6t_0$ after the start of the collapse in the parent simulation). The local angular momentum due to random convective motions in the envelope of our non-rotating star causes deflection from spherical infall as material reaches small radii. An outflow of $U \gtrsim GM/r$ and $\machr \sim 2-3$ material is launched.  The outflow drives a shock into the surrounding star.  The shock expands over time as accretion continues to feed the outflow with mass and energy.}
\label{fig:be_ma_timeseries}
\end{figure*}

\begin{figure*}
\centering
	\includegraphics[width=0.95\textwidth]{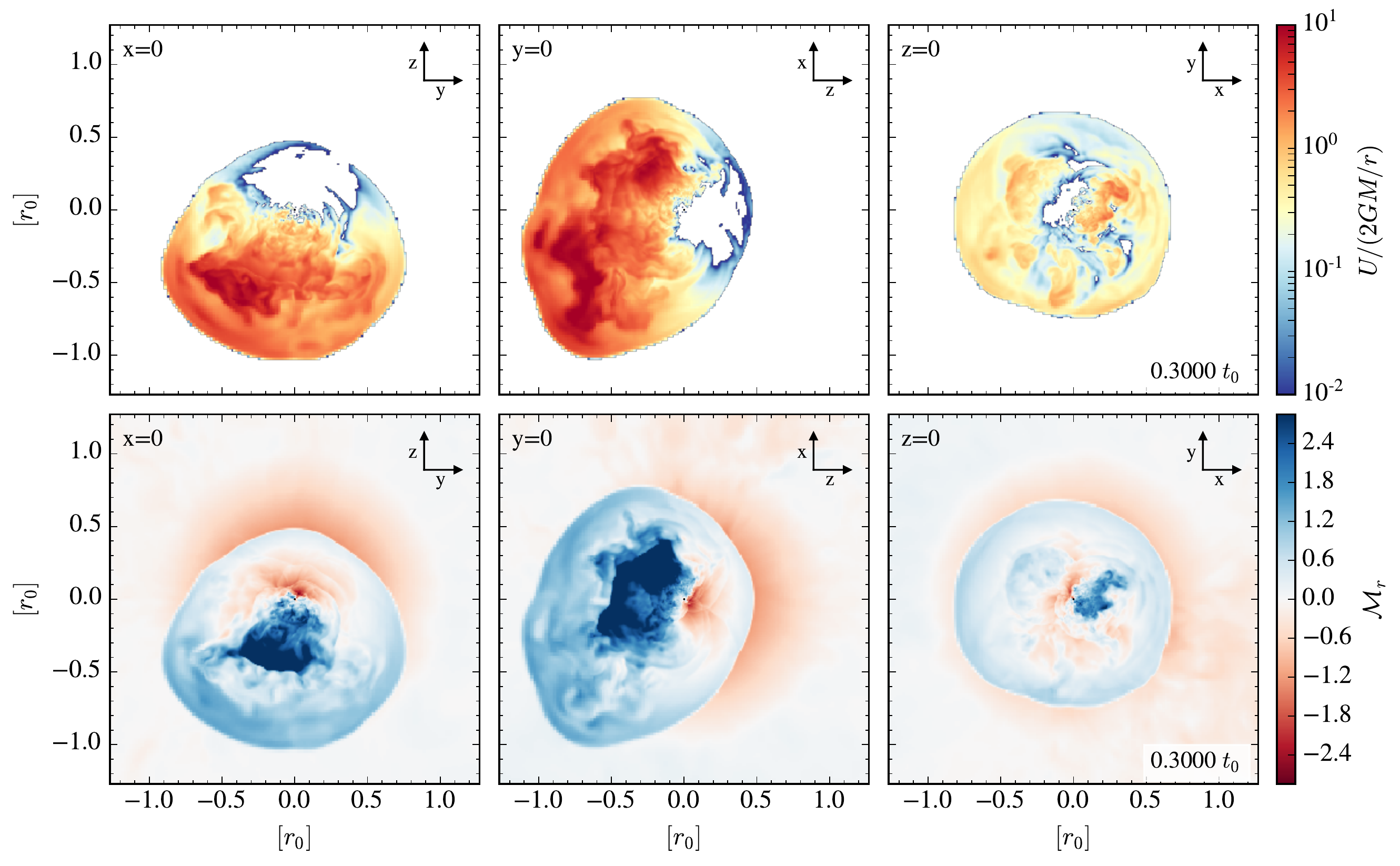}
    \caption{Same as in Fig.~\ref{fig:be_ma_timeseries} expect that we show slices through the three Cartesian planes at fixed time ($t =0.3 t_0$). The outflow is not very collimated and occupies a wide range of solid angle as it expands out through the stellar envelope. Interior to the outer shock, the first two columns show a small region of bound ($U <0$; white) inflowing gas at positive $z$ whose accretion feeds the outflow.}
\label{fig:be_ma_3planes}
\end{figure*}

Figures \ref{fig:be_ma_timeseries} and \ref{fig:be_ma_3planes} show slices of the gas flows at large scales in Model M.  The full simulation domain extends to $\pm 2.5 r_0$ but each panel of these figures is restricted to $\pm 1.1 r_0$ in the planes shown.   In each figure, the top row plots the local specific energy, 
\beq
U = \frac{v^2}{2} + \frac{P/\rho}{\gamma -1}  - \frac{GM}{r}
\label{eq:specificenergy}
\eeq
normalized by the square of the local escape velocity, $v_{\rm esc}^2 = 2GM/r$. In these panels, material with $U >0$ is plotted according to the log-scaled colourbar while material with $U < 0$ is not shown (white areas of the plot).  The bottom row in both figures plots the radial Mach number of the flow
\beq
\mathcal{M}_r \equiv v_r / c_{\rm s, ad}
\eeq
where $c^2_{\rm s, ad} = \gamma k_B T/\mu m_p$ is the adiabatic sound speed. We note that the limits of the (linear-scaled) colourbar are set to highlight the Mach numbers of material inside the expanding outflow.  The convective background has Mach numbers of only $\lesssim 0.3$ so is barely visible in these slices.  

Fig. \ref{fig:be_ma_timeseries} shows slices through the $y=0$ plane at three different times. The most striking large-scale feature is the bubble of unbound material that grows in time.  This is a consequence of energy and outflows generated by circularization at small radii, which drives a modest Mach number shock out through the infalling star.   Although the star is not rotating and has no net angular momentum,  the local angular momentum of the convective material reaching small scales breaks spherical symmetry, setting up an outflow of material with positive energy that expands outwards as time progresses (no such $U > 0$ outflow was present in the $\rsink > \rcirc$ simulations of Paper I).  The specific energy is only moderately greater than $2GM/r$, so continued accretion is required to drive the bubble to large radii and (as we will see in Sec.~\ref{sec:outflow}) unbind the star. Material within the bubble achieves (local) radial Mach numbers of 2-3.  We will show in Sec.~\ref{sec:outflow} that the velocity of the outer shock front is $\sim$ constant in time and radius.  Note also that, not surprisingly, the sustained accretion with $\machr <0$ inside the bubble is well-correlated with the bound gas (white color) in the specific energy panels. 

The three columns of Fig.~\ref{fig:be_ma_3planes} show (from left to right) slices through the $x=$, $y=$, and $z=0$ planes at fixed time. We note that the choice of these planes is for ease of plotting.  The Cartesian axes are arbitrary with respect to the flow as there is no preferred angular momentum axis in the envelope of our non-rotating star. The figure shows that the outflow is not collimated. Instead, the outer shock front occupies the full $4\pi$ in solid angle and all of the background material must cross the outer accretion shock, thermalizing some of the accretion energy. Because the outgoing shock is aspherical, it also deflects the incoming gas, increasing the dispersion in angular momentum inside the bubble.   Although the outer accretion shock occupies all solid angles, the outflow is not perfectly spherical and within the outflow there is spatial variation in energy and Mach number.  The specific energy and radial Mach number are weakest in the $z=0$ plane (third column).  In the other two planes (the first two columns), a large region can be seen that has the largest specific energy (dark red region in the upper panels) and largest positive radial Mach numbers (dark blue region in the lower panels).  Much of the $U < 0$ material accreting to small radii is being fed through a smaller range of solid angles (white region bordered by blue in the top row and dark red in the bottom row) that avoids running into the strongest outflow.  Although the flow structures at the large scales shown here and in Fig.~\ref{fig:be_ma_timeseries} do not change much as the bubble expands (for the most part, the earliest outflowing material is not overtaken by outflows launched at later times), the flow at small scales is much more time variable. 

\begin{figure*}
\centering
	\includegraphics[width=0.9\textwidth]{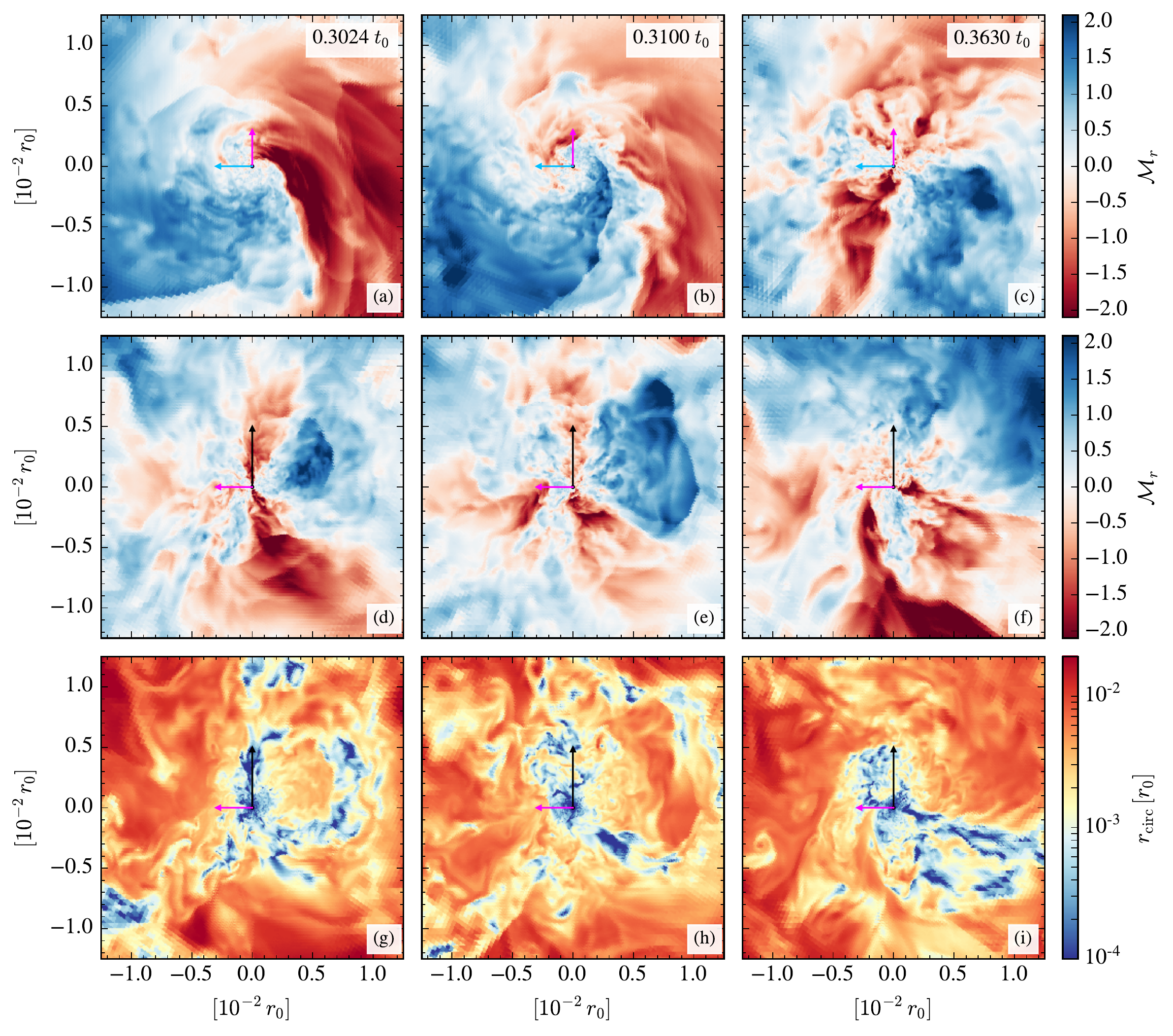}
    \caption{Slices of radial Mach number (top two rows) and circularization radius (bottom row) at three different times for Model M, highlighting the flow at small scales.  The slices are oriented such that the average angular momentum vector at $t=0.3024t_0$ (direction of the black arrow) points out of the page in the top row and lies within the page in the last two rows. The blue arrow shown in the top row points out of the page in the second and third rows. The pink arrow (orthogonal to the other two) is included for reference. Top row: Material with finite angular momentum circularizes in the plane of panel (a) and bypasses the sink. The outflowing material, in turn, moderately shocks and impacts the circularizing material flowing in at a somewhat later time in panel (b). By panel (c), the circularizing flow has been completely disrupted, replaced by radial streams of inflowing material; the mean angular momentum vector in this region is now perpendicular to a different plane. Middle row: While angular momentum support impedes the inflow in panels (a) and (b), streams of accreting material intersect the outflow in the perpendicular plane of panels (d) and (e), both along the angular momentum axis and from a third, random direction. There is no evidence of a disk in this ``edge-on'' view (nor at any other time) and outflows are seen to disrupt the inflow in this plane as well.  Bottom row: while much of the accreting material has $\rcirc < \rsink$, not all of the low-angular-momentum material is able to reach the sink and is instead advected away by the dominate outflow (e.g. blue material in upper, right side of panels g and h). Flow at these scales exhibit random inflow and outflow channels that change in time, but this time variability averages out at large scales, where the morphology remains mostly fixed as the outflow expands (see Fig.~\ref{fig:be_ma_timeseries}).}
\label{fig:mach_small}
\end{figure*}

Fig.~\ref{fig:mach_small} highlights the flow at small scales.  Each column in the figure shows the flow at fixed time (as labelled in the top panels).  The first and second rows plot radial Mach number while the third row plots the circularization radius of the material (here, $\rcirc = j^2 / GM$ where $j$ is the magnitude of the specific angular momentum vector in the zone). The (log-scaled) colourbar of the last row has a minimum of $\rsink$.  So, assuming each parcel conserves its specific angular momentum as time goes on, the dark blue material has low enough angular momentum to accrete. Everything else has $\rcirc > \rsink$.   Rather than plotting the Cartesian planes of the simulation as in previous figures, we have oriented the slices based on the direction of the average angular momentum vector of the material in this region, $\hat{\bmath\jmath}$, at the time shown in the first column ($t = 0.3024 t_0$).  The top row is the plane perpendicular to $\hat{\jmath}$ with $\hat{\jmath}$ pointing out of the page (i.e. if there were a disk, which there is not, this would be the ``face-on'' view).  The second and third row both show the same plane, whose normal vector, $\hat{n}$, is orthogonal to $\hat{\jmath}$ and points out of the page (i.e. if there were a disk, this would be the ``edge-on'' view). 

Fig.~\ref{fig:mach_small} (a) shows deflection of the inflowing material due to its finite angular momentum. The inflow cannot cool and we do not have a source of viscosity, so the infalling gas with $\rcirc > \rsink$ bypasses the sink and is launched back out with $\machr > 0$. Somewhat later, panel (b), the outflow interacts with and somewhat disrupts the swath of $\machr < 0$ material flowing in from the right side of the panel.  In the orthogonal plane, panels (d) and (e), material is able to reach the sink through narrow, supersonic streams that intersect the outflow and, in some cases, flow along the instantaneous angular momentum axis.  Comparing the radial Mach number plots in the second row to the circularization radius plots in the third row, we see that while it is the low-angular momentum material that often reaches the sink, not all of the gas that should be available to accrete does so. Instead, much of the low-angular momentum material (dark blue material in the third row) is impeded by or advected with the outflow.  Because the vast majority of gas has $\rcirc \gg \rsink$, the outflow dominates the dynamics.	We will see in Sec.~\ref{sec:accretion} that the mass accretion rate is suppressed relative to what is expected if one assumes that all of the material with $\rcirc < \rsink$ accretes.

There is no evidence of a disk in Fig~\ref{fig:mach_small} (top row) nor at any other time in the simulation. Even the circularizing flows of panels (a) and (b) and the somewhat collimated flows of panels (d) and (e) are messy, carrying the imprint of the large distribution of local angular momentum vectors in the flow.  The feeding direction of material falling in from larger radii is also time-variable. By the final time depicted, panel (c) shows no circularization of the flow in this plane, replaced instead by two radial streams of material feeding the sink.  The mean angular momentum vector has changed direction and any circularization is now happening in a different (random) plane.  We emphasize that while the directions of inflow and outflow at these scales is time-variable, these changes in direction average out at large radii.  Referring back to Figs.~\ref{fig:be_ma_timeseries} and \ref{fig:be_ma_3planes}, the outer shock and outflow simply expand without changing direction, even as energy is being fed into the bubble at smaller radii with random, time-variable $j$ and outflow directions. 

\begin{figure}
\centering
	\includegraphics[width=0.75\columnwidth]{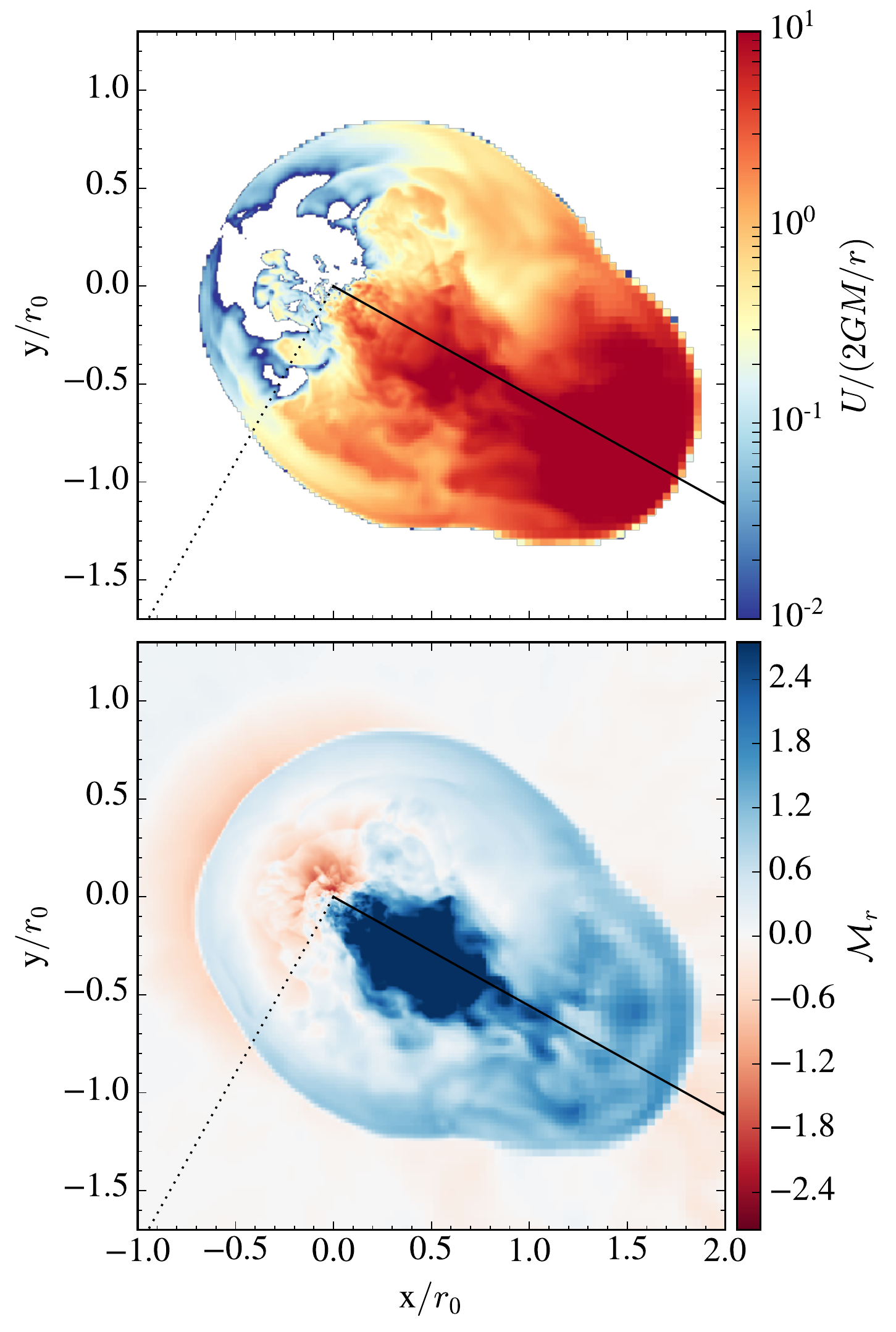}
    \caption{Slices through the $z=0$ plane of specific energy (top row) and radial Mach number (bottom row) for Model S at the end of the simulation ($t = 0.4t_0$). With a smaller sink radius and extra level of refinement around the sink, the flow is less spherical than shown in Figs.~\ref{fig:be_ma_timeseries} and  \ref{fig:be_ma_3planes} for Model M. The outflow morphology now has a slower-moving quasi-spherical component and a faster-moving plume of material with large $U$.  In addition, there is a large region of bound material within the more spherical shock region that continues to feed the BH. The dotted and solid lines are two lines of sight along which we estimate the shock velocity to be $1.8$ and $5.0$ $v_0$, respectively (see Sec.~\ref{sec:outflow}).}
\label{fig:be_ma_rs5em5}
\end{figure}

At the largest scales, the outflow of Model M is largely spherical (though it is non-spherical in key ways to facilitate long-term accretion).  We note, however, that the morphology of the outflow is sensitive to the size of the accreting sink and resolution.  Fig.~\ref{fig:be_ma_rs5em5} shows a snapshot of Model S in which $\rsink$ is a factor of 2 smaller and there is one extra level of refinement inside $r = 0.0003 r_0$ compared to Model M.  Model S has a more complex velocity structure than Model M, characterized by a slower-moving spherical component and a faster-moving plume. There is a factor of $\gtrsim 2$ difference in the shock velocity of these two components. 

We now consider quantitative results from our simulations, starting with accretion in the next subsection and then turning to the energy of the explosion in the following subsection. 

\subsection{Accretion}
\label{sec:accretion}
\begin{figure}

\centering
	\includegraphics[width=0.98\columnwidth]{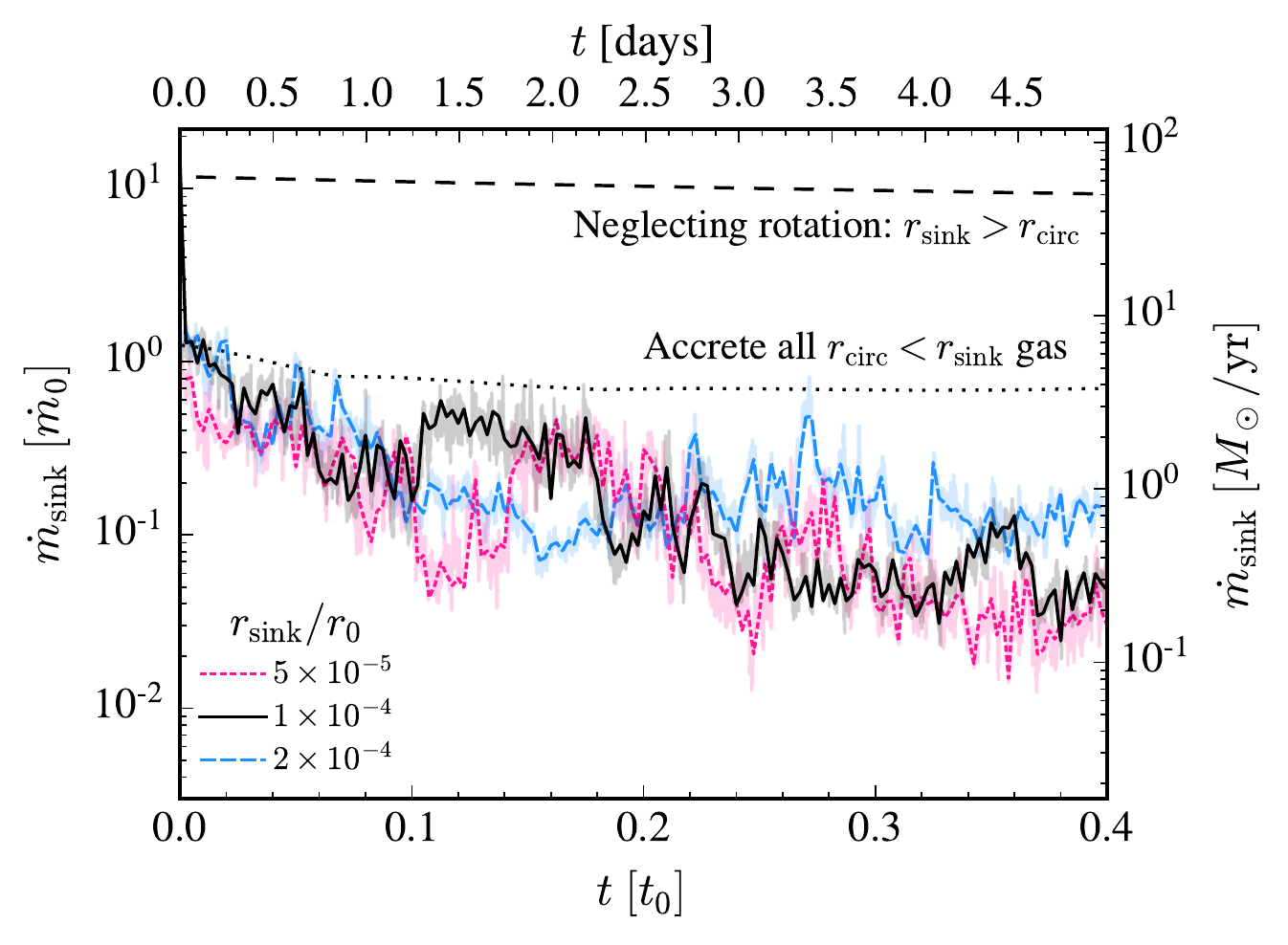}
    \caption{Instantaneous accretion rate versus time of matter falling into the absorbing sink of radius $\rsink$ for Models M (black, solid curve), S (pink, dotted curve), and L (blue, dashed curve). The values of $\rsink$ for each model are given in the legend.  The foreground, dark lines are the result of applying a spline to the raw data (transparent plots in the background), which permits readability when plotting the overlapping results of these models. The accretion rates are relatively independent of sink size. The accreted mass is similar for Models M and L.  For Models M and S, the accreted mass scales as $\propto \rsink^{0.6}$. All curves begin near the dotted line, which is the expected accretion rate assuming all gas with $\rcirc < 10^{-4} r_0$ initially is able to accrete.  The accretion rates fall further over time as the outflow impedes accretion of some of this low angular momentum gas. There is a factor of $\sim 100$ overall suppression of $\mdotsink$ relative to the infall rate when the star has no angular momentum (upper black, dashed line).}
\label{fig:mdot_vs_time}
\end{figure}

 In Fig.~\ref{fig:mdot_vs_time}, we plot the instantaneous accretion rate of gas entering the sink, denoted $\dot{m}_{\rm sink}$, versus time for our three high-resolution simulations.  For readability and comparison between models, the dark lines plot the data after a spline has been applied and the semi-transparent curves in the background show the raw, instantaneous data.  The black, solid curve corresponds to our fiducial model (M) while the pink, dotted and blue, dashed curves show models with smaller and larger sink radius, respectively, as noted in the legend. The left and bottom axes are in code units with time measured from the start of the zoom-in simulation.  The upper and right axes are converted to physical units applicable to a \mesa RSG model using Table~\ref{tab:units}.  The accretion rates are similar across simulations with varying sink size, dropping to roughly the same level over time and showing similar time-variability.  More quantitatively, we find that the total accreted mass is $0.102$, $0.101$, and $0.067$ $m_0$, respectively, for $\rsink = $2 , 1, and 0.5 $ \times 10^{-4} r_0$ (in our model RSG units, $m_0 \approx 0.2 \msun$, so roughly $2 \times 10^{-2} \msun$ is accreted during these simulations).

To interpret the overall normalization of these accretion rates, we include two additional curves in Fig.~\ref{fig:mdot_vs_time}.  The black, dashed curve shows the accretion rate realized in a simulation with $\rsink > \avercirc$ (that is, if all of the matter in each shell accretes without feedback). The dotted line is the accretion rate one would expect for the present simulation if only the material in each shell with $\rcirc < 10^{-4}r_0 \approx \rsink$ prior to collapse were able to accrete.  Initially, $\mdotsink$ for all models is consistent with accretion of all of the low-angular-momentum material with $\rcirc < \rsink$  (the curves begin near the dotted line). As we saw in Fig.~\ref{fig:hist}, most of the material has $\rcirc \gg \rsink$ so cannot accrete, instead driving an outflow. Once the outflow begins, the accretion rate drops by a factor of $\sim$ten.  There are two main reasons for this difference between the black, dotted line and the simulated accretion rates. First, the outflow rearranges the local $j$ of all the gas, so $\mdotsink$ at later times cannot be predicted from the initial state of the gas. Even more importantly, as was apparent in Fig.~\ref{fig:mach_small}, low-angular momentum material that would otherwise be able to accrete is deflected by the outflow, so the accretion rate is suppressed even further than a naive prediction based on the initial $\rcirc$ of the material alone. Overall, the angular momentum associated with convection results in a factor of $\sim$100 suppression in accretion relative to the collapse of a RSG neglecting angular momentum (dashed line).

\begin{figure*}
\centering
	\includegraphics[width=0.85\textwidth]{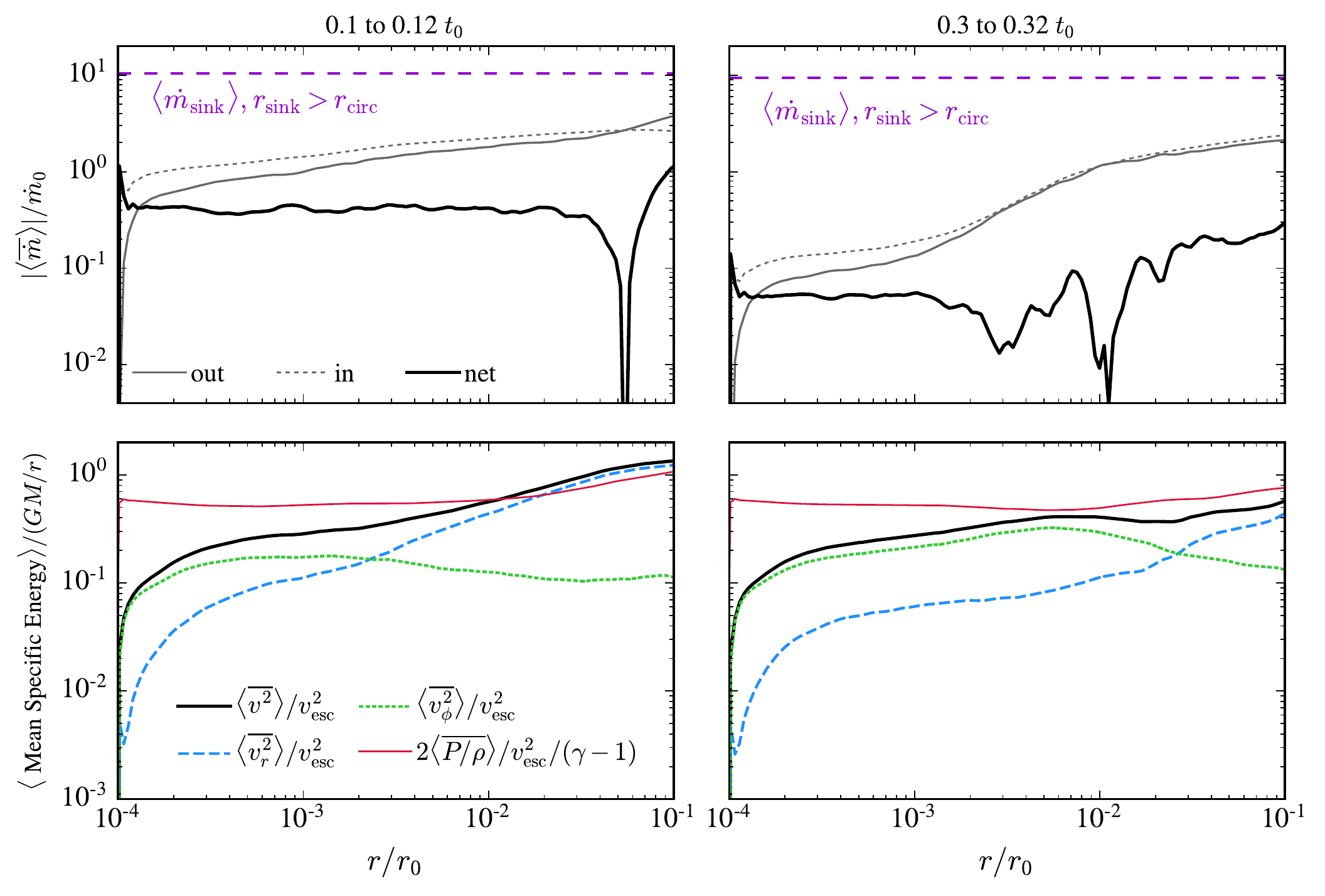}
    \caption{Time- and spherically-averaged radial profiles of the absolute value of the mass accretion rates (top row) and specific energies (bottom row) for Model M with time averages computed over 0.1 to 0.12 $t_0$ (left column) and 0.3 to 0.32 $t_0$ (right column). Top row: The net accretion rate is a factor of 10-100 lower than the accretion rate expected for a star without angular momentum (purple dashed line) as a similar rate of mass flows out (thin, solid line) as flows in (thin, dotted line). At both times, $\dot{m}_{\rm in}(r) \propto r^{0.2}$ for $\rsink < r < \rcirc$.  At later times, the accretion rate decreases as the gas becomes more angular-momentum supported.  Bottom row: Angular momentum support (green dotted line) is always subdominant to thermal pressure support (solid red line) but still dominates the energy density in radial flows (blue, dashed line) at small radii. At later times, angular momentum support is higher and extends to larger radii as much of the low-angular momentum material has already accreted.  }
\label{fig:time_ave_profiles}
\end{figure*}

Critically, Fig.~\ref{fig:mdot_vs_time} also shows that $\mdotsink$ never drops to zero (even after $t \sim 0.14t_0$, when the total energy of the star in models M and S turns positive; see Sec.~\ref{sec:outflow}).  As we showed in Fig.~\ref{fig:mach_small}, material reaches small radii through narrow streams that intersect the rising plumes.  The top row of Fig.~\ref{fig:time_ave_profiles} quantifies these roughly co-spatial inflows and outflows, showing the time-averaged radial mass flux profiles, $\langle \dot{m}(r) \rangle$, at two different times in Model M where 
\beq
\dot{m}(r) = 4\pi r^2\overline{\rho v_r}(r)
\eeq
is the instantaneous, spherically-averaged mass flux.   In each panel, the dotted, grey line is the absolute value of the inward mass flux (material with $v_r < 0$), while the thin, grey line is the outward mass flux (material with $v_r > 0$). The thick, black line is the absolute value of the net mass flux.  At both times, the net mass flux has roughly equal contributions between inflow and outflow with inflow being somewhat larger than outflow inside of $\sim 0.1 r_0$.  The potential energy liberated as the inflow reaches small radii allows continued feeding of the energetic outflow.  

The lower panels of Fig.~\ref{fig:time_ave_profiles} compare different sources of (time- and spherically-averaged) specific energy in the inner region of the simulation at the two times shown in the top row. We have normalized each specific energy term by $v_{\rm esc}^2 = 2GM/r$.  The black, solid curve is the total specific kinetic energy while the blue, dashed curve is the contribution to the specific kinetic energy from radial motion only.  The solid, red curve is the thermal energy.  Our flows have randomly oriented angular momenta without a fixed rotation axis in space or in time.  To associate a specific kinetic energy with centrifugal support without reference to a particular rotational axis, we measure the magnitude of the angular momentum of each cell, $j$, then locally define $v_\phi \equiv j/r$.  The kinetic energy associated with rotational support in each cell is $v_\phi^2/2$.  The green, dotted line in the two lower panels is the time- and spherical-average of $v_\phi^2$.	
		
Fig.~\ref{fig:time_ave_profiles} shows the mean flows are sub-Keplerian and thermal pressure is always dominant.  At small radii, the rotational energy is larger than that associated with the radial flows.  The amount of rotational support increases somewhat and extends to larger radii as time progresses.  Over time, more and more low-angular momentum material has been accreted and there is also mixing and sharing of angular momentum. The result is increased rotational support over time in the inner regions.  We emphasize again, however, that there is not a fixed axis for the angular momentum vector and therefore no persistent disk. 

\subsection{Energetics of the explosion}
\label{sec:outflow}
We have seen that the finite angular momentum arising from random convective motions is sufficient to launch an outflow and suppress the accretion rate by a factor of $\sim100$ relative to the zero-angular-momentum case.  In this subsection we will quantify the energy and shock speed of the explosion driven by that outflow. 

\begin{figure}
\centering
	\includegraphics[width=0.98\columnwidth]{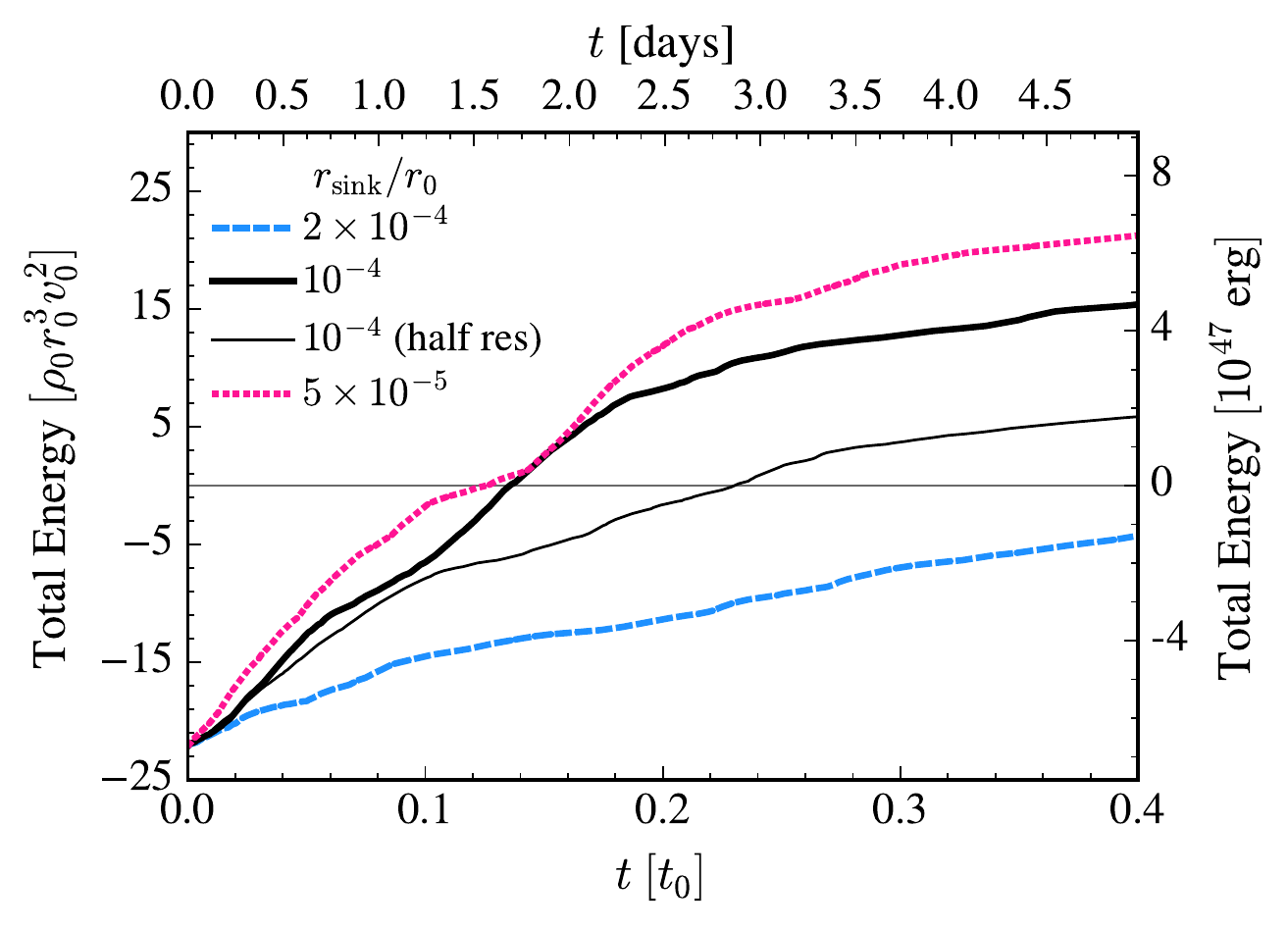}
    \caption{Integrated total energy of the explosion over time for Models M (thick, black line), S (pink, dotted line), L (blue, dashed line), and \Mhalf (thin, black line). The top and right-hand axes scale code units to a 16.5$\msun$ (at collapse) \mesa RSG using Table~\ref{tab:units}. Smaller sink radius and higher resolution each serve to increase the total energy of the explosion. The star is unbound when the curves become positive.  Critically, the energy continues to increase after this point as accretion continues and the outflow continues to be fed with mass and energy.}
\label{fig:bubble_energy}
\end{figure}

Fig. \ref{fig:bubble_energy} plots the total energy of our star over time for our fiducial model (Model M; thick, solid line), our simulation with smaller sink (Model S; pink, dotted line), our simulation with larger sink (Model L; blue, dashed line), and the half-resolution simulation that is otherwise identical to Model M (Model \Mhalfns; thin, solid line).  The sink radius of each model is given in the legend. At each time, we integrate the total energy of the star out to a radius of $8r_0$ (exterior to our model photosphere) by using both the zoom-in simulation and the parent simulation at the corresponding time.  The total energy in each grid zone is 
\beq
d\mathcal{E}_{\rm tot} = \rho U dV,
\label{eq:total_E_zone}
\eeq
where $dV$ is the volume of the zone\footnote{Note that we do not have hydrogen recombination energy in our equation of state, so it is not included in this energy budget.}.  In the zoom-in simulation, we integrate eq. \eqref{eq:total_E_zone} out to a radius of $r = 2.25 r_0$, which is larger than the size of the outgoing shock at all times in the simulations shown. In the parent simulation, we integrate eq. \eqref{eq:total_E_zone} from $2.25 r_0$ to $8 r_0$.  The sum of these two integrals, $\mathcal{E}_{\rm tot}(t)$, is the total energy plotted in Fig.~\ref{fig:bubble_energy}.  Our fiducial model crosses zero total energy at $\sim 0.13 t_0$. Beyond this point, the star has net positive energy and an  unbound outflow.  At the end of the simulation, the outflow in our fiducial model has a net energy of $\sim 5 \times 10^{47}$ erg when scaled to a $16.5 \msun$ RSG.  Note, however, that the excess energy is not uniformly distributed over the entire $4\pi$ of material. A small amount of mass is still bound and accretes via narrow lanes that penetrate the unbound outflow.  The total energy continues to grow beyond $t \sim 0.13r_0$ because the inflow continues to feed the outflow with both mass and energy. The simulations shown in Fig. \ref{fig:bubble_energy} end before unbound material reaches the surface of the star because the domain extends only to $r \sim 2.5 r_0$ while the photosphere is located at $\sim 6r_0$.  We will show in Sect.~\ref{sec:tparent}, using a simulation at half-resolution, that we expect the total energy to increase monotonically until the shock reaches the surface of the star. Thus, the true explosion energy would be larger than shown here in Fig.~\ref{fig:bubble_energy}. 

The three additional curves in Fig.~\ref{fig:bubble_energy} show how the total energy depends on sink radius and resolution.  The blue, dashed and pink, dotted curves show how the total energy changes when changing the sink radius by a factor of 2.  Decreasing the sink size systematically increases the total energy of the explosion because the material can reach smaller radii and liberate accretion energy deeper in the potential well.  Note that the first halving of the sink radius from $2 \times 10^{-4}$ to $1 \times 10^{-4}r_0$ results in a greater energy increase than the next halving in sink radius (to $5 \times 10^{-5} r_0$). It is not clear at what sink radius the simulations would converge, but it is likely the increase in energy is becoming less pronounced because $r_{\rm sink} = 10^{-4}r_0$ is already smaller than the initial circularization radius of most of the material.  Including magnetic fields or another form of angular momentum transport in our simulations would likely produce a different dependence on $\rcirc$ as it would allow the material to shed more of its initial angular momentum, increasing the mass flux to small radii and the energy supplied to outflows.   Finally, the thin, solid curve for Model \Mhalf shows that reducing the base resolution by a factor of 2 reduces the final total energy by a factor of $\sim 2.6$ compared to Model M.  More physical simulations with smaller sink radius and higher resolution would, thus, serve to increase the total energy of the explosion in our hydrodynamical simulations. Therefore, the energies that we find here should be regarded as lower limits on the explosion energy.

We estimate the time-averaged accretion efficiency for the models plotted in Fig.~\ref{fig:bubble_energy} by comparing the change in total energy in the gas, $\Delta \mathcal{E}_{\rm tot} = \mathcal{E}_{\rm tot}(t_f) - \mathcal{E}_{\rm tot}(t_i)$, to the maximum available energy from accretion, $\mathcal{E}_{\rm acc} \equiv \int_{t_i}^{t_f} (GM\mdotsink/\rsink) dt$. We find that the time-averaged efficiency, $\Delta \mathcal{E}_{\rm tot}/\mathcal{E}_{\rm acc}$, over roughly the duration of the simulation is $3-5\%$, with efficiency increasing with resolution. This efficiency is a factor of $\sim$few-$10$ less than what one might naively expect for a rotating flow. The reason is that the accreted gas is weakly bound with Bernoulli parameters of $\sim -0.1 GM/\rsink$.

\begin{figure}
\centering
	\includegraphics[width=0.98\columnwidth]{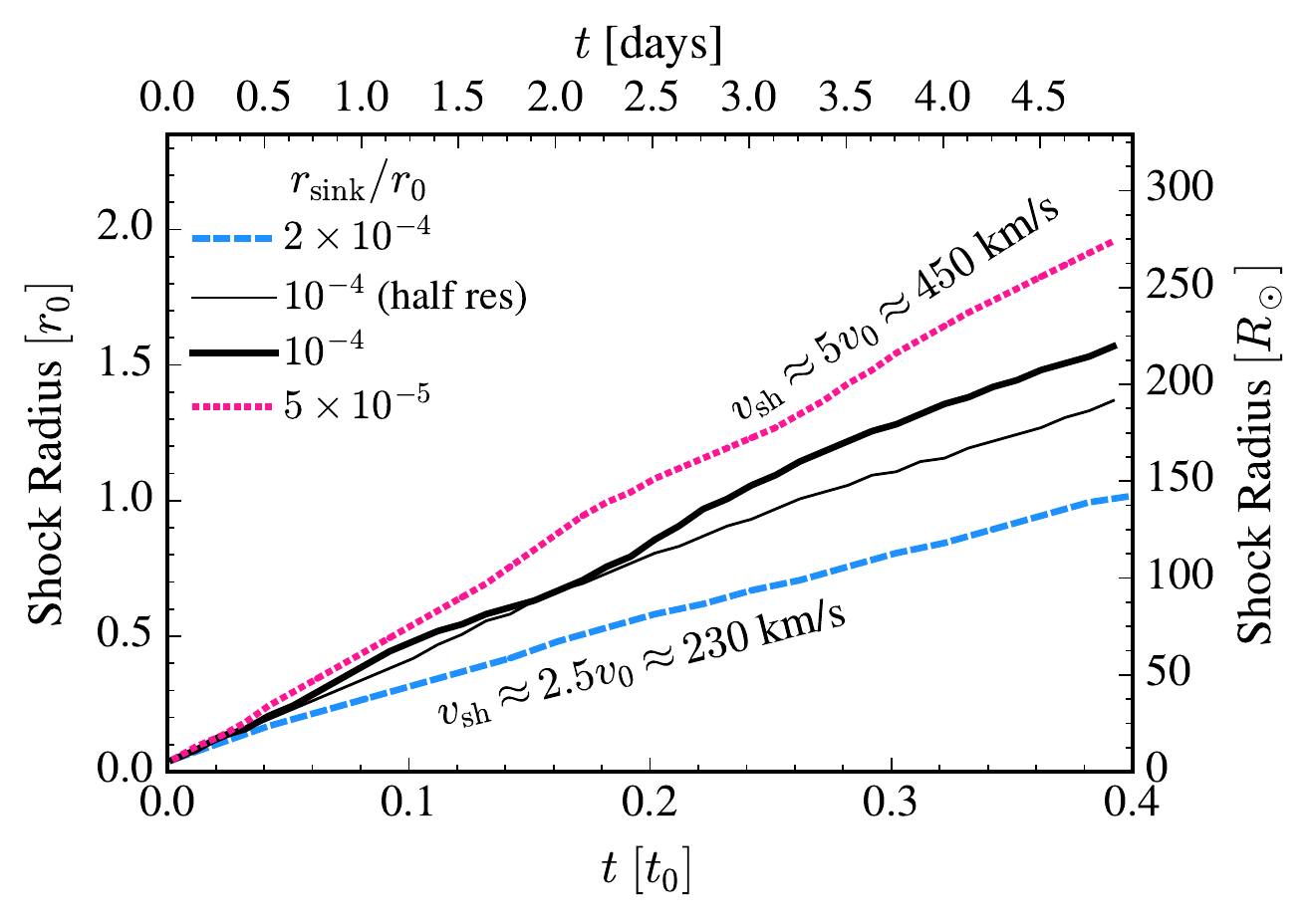}
    \caption{Radius of the leading edge of the outflow as a function of time for the models shown in Fig.~\ref{fig:bubble_energy}. The upper and right-hand axes scale code units to the same \mesa RSG model as in that figure. Our fiducial model (thick black line) has an average speed of $v_{\rm sh} \approx 4 v_0$ or $\approx 360$ km/s in units of the RSG model.}
\label{fig:shock_radius}
\end{figure}

We showed in Figs.~\ref{fig:be_ma_timeseries} and \ref{fig:be_ma_3planes} that a ``bubble'' of large specific energy and relatively high Mach number material is inflated as ongoing accretion feeds the outflow with mass and energy.  As the bubble expands, the surface of the bubble sweeps through and shocks the surrounding star. It is helpful to determine the radius of the shock, $r_{\rm sh}$, as a function of time.  Although our outflow is not perfectly spherical, we define a single value for $r_{\rm sh}$ by computing the spherically-averaged Bernoulli profile and setting $r_{\rm sh}$ to be the radius at which the Bernoulli profile changes sign from positive to negative.  In practice, this method picks out the outer-most radius occupied by the outflow, that is, the fastest moving point on the surface of the outflow.  In Fig.~\ref{fig:shock_radius}, we plot $r_{\rm sh}$ as a function of time for the same models shown in Fig.~\ref{fig:bubble_energy}.   The slope of each curve is the speed of the leading edge of the shock, $v_{\rm sh}$. The shock speed increases with decreasing sink size and decreases at lower resolution. The shock speed is relatively constant for the simulations shown, varying between 2.5 and 5 $v_0$. Our fiducial model (thick, solid line), has an average slope of $4 v_0$.  When scaled to the same \mesa model as in Fig.~\ref{fig:bubble_energy}, $v_0 = 90.6$ km/s, giving shock speeds of $\sim$ few hundred km/s. 

Model S, which has the smallest sink radius and highest resolution of our models, also has the most complex velocity structure. The dotted and solid lines of Fig.~\ref{fig:be_ma_rs5em5} show two lines of sight along which we estimate $v_{\rm sh}$ in the $z = 0$ plane. The fastest moving material (along the solid line-of-sight) has $v_{\rm sh} \approx 5v_0$, which is the shock speed picked out by our method above for computing the pink, dotted line in Fig.~\ref{fig:shock_radius}. Along a second line-of-sight that is 90 degrees away (dotted line of Fig.~\ref{fig:be_ma_rs5em5}), $v_{\rm sh} \approx 2v_0$. This more-complex velocity structure could alter the spectra for these events compared to a more spherical outflow.

Although our shock velocities are $\sim$constant, the expansion of the shock during our simulations is neither in the free-expansion nor Sedov-Taylor regime.  The shock is still interior to the star so is sweeping up significant mass, but it is not Sedov-Taylor because energy is constantly being injected into the outflow from ongoing accretion.  Indeed, the constant shock velocity is somewhat of a coincidence, which can be seen as follows.  Let 
\beq
E_{\rm sh} \propto m v_{\rm sh}^2 \propto t^n
\label{eq:twiddleE}
\eeq
be the energy of the shocked gas as a function of time with $m$ the mass swept up by the shock and $v_{\rm sh}$ the shock velocity.  The radius of the shock as a function of time is $r_{\rm sh} = v_{\rm sh} t$.  $E_{\rm sh}(t)$ and, thus, $n$ are set by the physics at small radii that is injecting energy into the shock.  Recall that the density profile of our polytropic envelope is $\rho \propto r^{-b}$.  The swept-up mass as a function of shock radius is then $m \propto \rho r_{\rm sh}^3 \propto r_{\rm sh}^{3-b}$. 
So $mv_{\rm sh}^2 \propto r_{\rm sh}^{3-b}(r_{\rm sh}/t)^2 \propto r_{\rm sh}^{5-b}t^{-2}$. From eq.\eqref{eq:twiddleE},   $r_{\rm sh}^{5-b}t^{-2} \propto t^n$ so 
\beq
r_{\rm sh} \propto t^{(n+2)/(5-b)}.
\eeq
For $b \approx 2$, as in our model (and RSG envelopes, in general), $r_{\rm sh} \propto t^{(n+2)/3}$, $n=1$ implies $r_{\rm sh} \propto t$, that is, a constant shock speed.   For modest variations in $n$, e.g. $n\sim 0.4 - 1.5$, $r_{\rm sh} \propto t^{0.8 - 1.16}$, consistent with Fig.~\ref{fig:shock_radius}. 

\subsection{Models initialized at the beginning of the collapse}
\label{sec:tparent}
In the previous subsections, we presented results from our models that were initialized from the $t_p = 0.6t_0$ parent snapshot.  That snapshot was chosen because the infalling material had a large circularization radius relative to earlier times in the parent simulation, thus permitting us to use a larger sink radius.  The longer time-step associated with a larger sink, in turn,  allowed us to study the flows at the circularization radius at high resolution and to follow the outgoing shock well beyond the point when the total energy of the star became positive.  

\begin{figure}
\centering
\includegraphics[width=0.98\columnwidth]{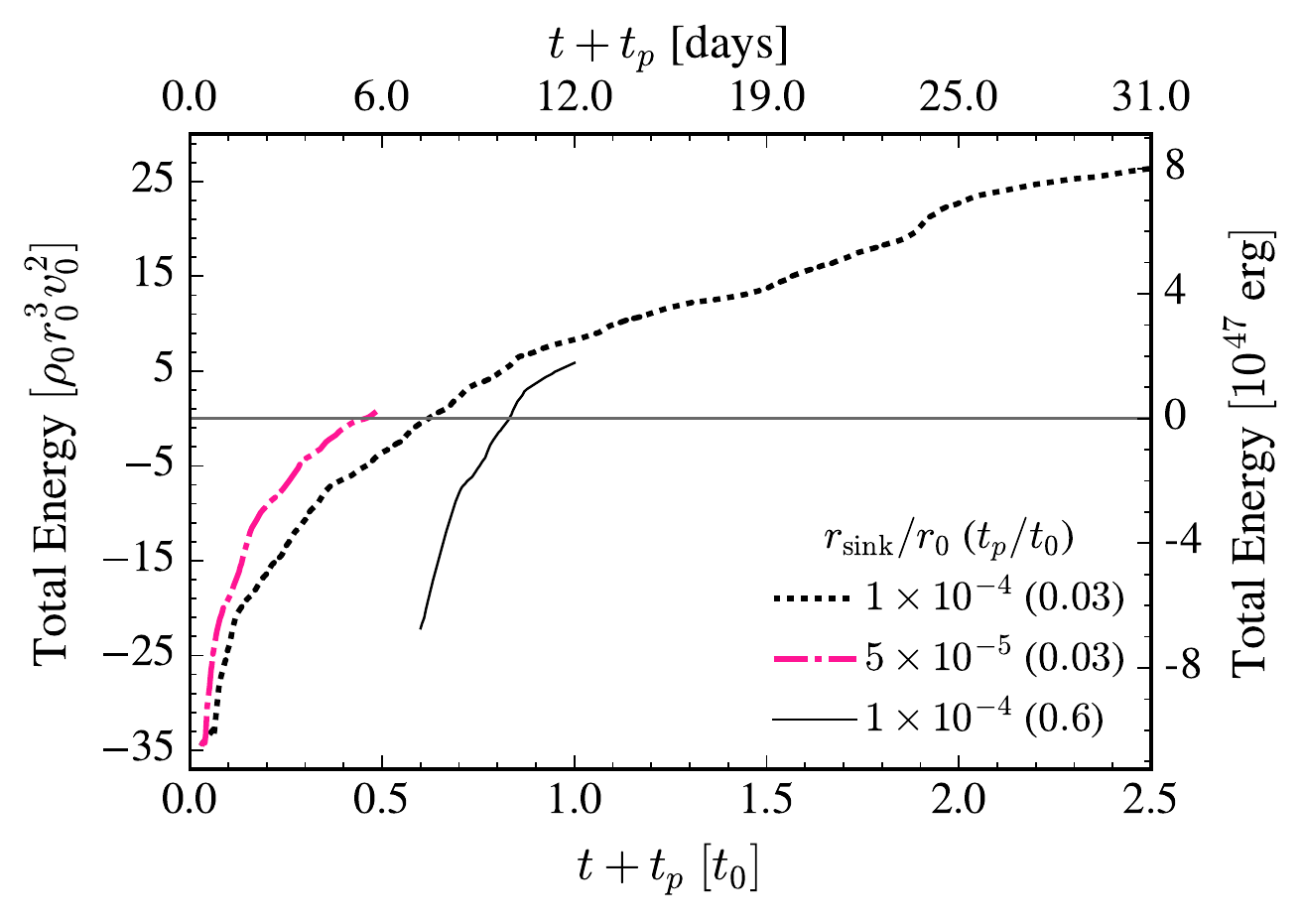}
    \caption{Total energy versus time for Models \Mhalftp (black, dotted), \Shalftp (pink, dash-dotted), and \Mhalf (thin, solid).  We have added $t_p$ to times in the zoom-in simulation so that the $x$-axis gives the elapsed time since the start of the collapse in the parent simulation. Comparing \Mhalftp to \Mhalf at $t + t_p \approx 1.0t_0$, the two simulations give similar total energies.  Comparing the black, dotted and pink, dash-dotted curves, the explosion energy indeed increases with smaller sink, as in Fig.~\ref{fig:bubble_energy}. Initializing models from earlier $t_p$ yields only minor differences in our results. When scaling code units to our reference \mesa model, Model \Mhalftp reaches $8\times 10^{47}$ erg at the end of the simulation.} 
\label{fig:tp_energy}
\end{figure}

In this section, we present the results of Models \Mhalftp and \Shalftpns, which are initialized, instead, from $t_p = 0.03t_0$ (this is equivalent to initializing from the true start of the collapse in the parent simulation at $t_p=0$; Sec.~\ref{sec:refinement_and_parameters}).  Model \Mhalftp is a long-running simulation with a larger domain at half resolution which models closer to the full collapse and explosion of the star compared to Model M. The \Shalftp model has a factor of 2 smaller sink radius and is used to confirm that the behavior of the results with sink radius still holds with $t_p=0.03t_0$ and at half resolution.  Details of the grid structure are given in Sec.~\ref{sec:refinement_and_parameters}.  In the remainder of this subsection, we show how initializing from an earlier parent snapshot leads only to minor differences in our results.

Fig.~\ref{fig:tp_energy} shows the total energy of the star for Models \Mhalftp (dotted line), \Shalftp (pink, dash-dotted line), and \Mhalf (solid line) as a function of time since the start of collapse in the parent simulation, $t + t_p$.  The total energy for Models \Mhalftp and \Shalftp is computed by integrating eq.~\eqref{eq:total_E_zone} to $r=6.0r_0$ in the zoom-in simulation and over $6.0 < r/r_0 < 8.0$ in the parent simulation at the equivalent time.  The curve for Model \Mhalf is the same as in Fig.~\ref{fig:bubble_energy} but is repeated here to isolate the impact of varying $t_p$ alone.  The right-hand and upper axes are scaled to physical units as in Fig.~\ref{fig:bubble_energy}.   The curves with $t_p = 0.03t_0$ begin at lower total energy because the innermost part of the envelope with the highest density and binding energy is still on the domain.  In the $t_p = 0.6t_0$ snapshot used to initialize all other models, on the other hand, the outgoing rarefaction wave that began at $t_p = 0$ has already reached $r \approx  0.7 r_0$ and the envelope inside of $r\approx 0.4$ $r_0$ has already accreted.  The initial slope of the \Mhalf curve is somewhat steeper because of these different initial conditions. However, at $t + t_p = 1.0 t_0$ since the start of collapse, \Mhalftp and \Mhalf have very similar total energy.  Comparing the black, dotted and pink, dash-dotted lines, we also see an increase in energy in the outflow with smaller $\rsink$, just as we found for the models of Fig.~\ref{fig:bubble_energy}.   Though not shown, we also find that $r_{\rm sh}(t)$ for the $t_p = 0.03 t_0$ models are similar to that of model \Mhalfns.  The shock velocities are $\sim$constant, with $v_{\rm sh} \approx 2.3 v_0$ in model \Shalftp and $v_{\rm sh} = 1.5-2.0 v_0$ in the \Mhalftp model. We conclude that our simulation results do not depend very strongly on $t_p$, i.e., on when in the original collapse simulation we begin our zoom-in simulations. 

\section{Extrapolation to real RSG collapse}
\label{sec:extrap}
In this section, we first estimate the explosion energy we would obtain if we ran a simulation until breakout of the shock. We then speculate as to how further reductions in sink size or inclusion of magnetic fields might impact the explosion energy obtained in simulations.

Model \Mhalftp was run until $t = 2.5t_0$. At this time, $r_{\rm sh} \approx 4.0r_0$ and $v_{\rm sh} \approx 1.5 v_0$. The photosphere of the star is at $\sim 6 r_0$.  Assuming constant $v_{\rm sh}$, and neglecting shock acceleration near the surface of the star, the shock will reach $6.0r_0$ by a time of $\approx 3.7t_0$.  If we extrapolate the model \Mhalftp total energy curve (Fig.~\ref{fig:tp_energy}, dotted line) to $3.7t_0$, we find the model reaches a total energy of $\approx 10^{48}$ erg when scaled to our reference \mesa model.   Comparing the thin, black line and pink, dotted lines of Fig.~\ref{fig:bubble_energy}, the combined effects of a factor of 2 finer base resolution and a factor of 2 smaller $\rsink$ raises the outflow energy by a factor of $\approx 4$. Thus, if we could simulate at our highest resolution and smallest sink radius until the outflow reaches the surface of the star, the model would reach a total energy of $\sim 4 \times 10^{48}$ erg.  

We also measure the ejecta mass as a function of time (defined as the mass in the outflow with $U>0$; not shown). Extrapolating that curve to $3.7t_0$, we find that  $M_{\rm ej} \approx 10.5\msun$ for our reference \mesa model (recall from Fig.~\ref{fig:mdot_vs_time} that $\mdotsink$ is suppressed by a factor of $\sim 100$ due to the presence of the random angular momentum of convection, so the ejected mass is very nearly the total mass of the hydrogen envelope). Combining our estimates for the explosion energy and ejecta mass, the characteristic speed of the explosion would be $\approx \sqrt{2\mathcal{E}_{\rm tot}/M_{\rm ej}} \approx$ 200 km/s. 

Thus, within our simulation framework, collapse of the convective envelope of non-rotating RSGs would give rise to explosions of at least $\sim$ few $\times 10^{48}$ ergs and characteristic velocities of $\sim$ 100s km/s.   We emphasize that this is likely a lower limit.  First, the supply of energy into the explosion does not stop when the shock reaches the surface of the star.  This is simply a convenient time at which to estimate the minimum energy of the explosion.  Accretion will continue as there is still material interior to the shock with $U<0$ and $\machr < 0$.  Only a small amount of mass is required to feed the outflow (e.g. only a fraction of $\sim 10^{-2} \msun/10.5\msun \sim 10^{-3}$ of the hydrogen envelope is accreted by $2.5t_0$ in Model \Mhalfns).  Even if the vast majority of the hydrogen envelope is ejected, it would likely take another stellar dynamical time before enough expansion of the envelope has occurred to begin shutting off accretion at small radii. Additionally, our simulations suggest that the explosion energy increases as we decrease the size of the accreting sink, but we do not know at what sink radius the explosion energy will converge. Ideally, one would want to employ $\rsink = \risco$, which is a factor of $\sim 100$ decrease from what we have simulated here.  In studies of radiatively-inefficient accretion flows (RIAFs), the accretion rate has been found to scale as $\dot{m} \propto r^s$ with $0.5 \lesssim s \lesssim 1$ \citep{2005ApJ...628..368O,2011MNRAS.415.1228P,2020MNRAS.492.3272R,2022ApJ...934..132H}.  For $s = 1/2$, the accretion luminosity would scale as $\dot{E} \propto \dot{m}/r \propto r^{-1/2}$, leading to an increase in $\dot{E}$ of $\sim 10$ for a factor of 100 decrease in sink size.  On the other hand, such an energetic outflow may efficiently shut off accretion, so it is unclear whether all of this increase in $\dot{E}$ is realizable. 

The scalings found for RIAFs rely on a source of angular momentum transport to allow  the material to continuously shed angular momentum.  Magnetic fields are undoubtedly critical to this process, both via the magneto-rotational instability and large-scale magnetic stresses. Future simulations that include magnetic fields are critical for determining the range of explosion energies that may arise from accretion of the convective RSG envelope onto the newly formed BH following nominally failed SNe.

\section{Observational implications for RSG collapse}
\label{sec:obs}
We have shown that infall of convective material drives an accretion shock out through the collapsing star with enough energy to unbind the envelope.  After the accretion shock exits the stellar surface, the ejecta continue to expand due to the excess thermal energy deposited behind the shock.  As this material expands and cools, hydrogen recombination begins in the outermost layer. Recombination then proceeds inward in mass coordinates through the expanding ejecta. The location of the recombination front sets the location of the photosphere, giving rise to a plateau in the light curve. This is the same phenomenon as in SNe IIp, but the explosion energies are lower in our scenario, so the ejecta starts off cooler.  The energy released from hydrogen recombination is also more similar to the explosion energy than in SNe IIp.  

As discussed in the previous section, we have only a rough estimate for the minimum explosion energies that might be realized in nature due to the infall of the convective hydrogen envelope, so let us consider the luminosities and durations of the plateaus that might be achieved across a full range of energies. 
\citet{2010MNRAS.405.2113D} performed radiation hydrodynamical calculations of explosions of RSG envelopes covering explosion energies of $\sim 10^{48}$ to $\sim 10^{51}$ erg.  These models are particularly applicable to our scenario as they do not include radioactive heating (our explosions do not lead to nucleosynthesis) but do include realistic opacities and heating that capture the changing ionization state of the hydrogen as the material expands.  Fig. 8 of \citet{2010MNRAS.405.2113D} shows the important characteristic features of the light curves at the two energy extremes.   At high energies, the light curve manifests as a SNe IIp, but without a nickel tail following the plateau \citep[see also fig. 4 of][]{Goldberg2019}.  At low explosion energies, the luminosity on the plateau becomes brighter than the initial shock-cooling luminosity and the lower ejecta velocities extend the plateau duration to $\gtrsim$ year.  This can also be seen in \citet[][fig. 12]{2013ApJ...769..109L}, in which a $10^{47}$ erg explosion of a $\sim 15 M_\odot$ RSG results in a $\sim$ year-long plateau of $\sim$ few $\times 10^{39}$ erg/s. 

In what follows, we use the one-zone model of \citet{DK2013} to make rough estimates of the plateau luminosity, $\Lpl$, and duration of the light curve, $\tpl$, for a variety of explosion energies, ejecta masses, and initial stellar radii.   Appendix A of \citet{DK2013} computes the luminosity due to diffusion of the thermalized and adiabatically-degraded explosion energy out to the photosphere. Once the temperature of the material reaches the hydrogen recombination temperature (at time $t_i$ in their notation), they account for recession of the photosphere (coincident with the recombination front) through the expanding ejecta (see also \citealt{Popov1993}). Without radioactive heating ($H = 0$ in their model), the luminosity as a function of time, $L(t)$, is given by their equation (A13).\footnote{The factor $t_i^{6/5}t^{4/5}$ should be $t_i^{4/5}t^{6/5}$ in \citet{DK2013} eq. (A13).} The light curve ends when the photosphere reaches a radius of zero (at time $t_f$ in our notation).    To compute $\Lpl$, we evaluate $L(t)$ at $(t_i + t_f)$/2 which roughly coincides with the maximum of $L(t)$ on the plateau.  We take for the event duration $\tpl = t_f - t_i$, which is the time during which recombination sets the photosphere.  For these estimates, we adopt a hydrogen recombination temperature of 6000 K and assume the opacity interior to the photosphere is set by electron scattering.   The values of $\Lpl$ that we obtain are in good agreement with the simulated light curves of \citet{2010MNRAS.405.2113D}, but $\tpl$ is too long by a factor of $1-1.4$. This estimate is sufficient for our purposes here, but more careful calculations should account for recombination photons as an energy source, should relax the assumption of a strong shock, and should include the density and temperature profile behind the shock rather than the one-zone model employed here. 
\begin{figure*}
\centering
	\includegraphics[width=0.8\textwidth]{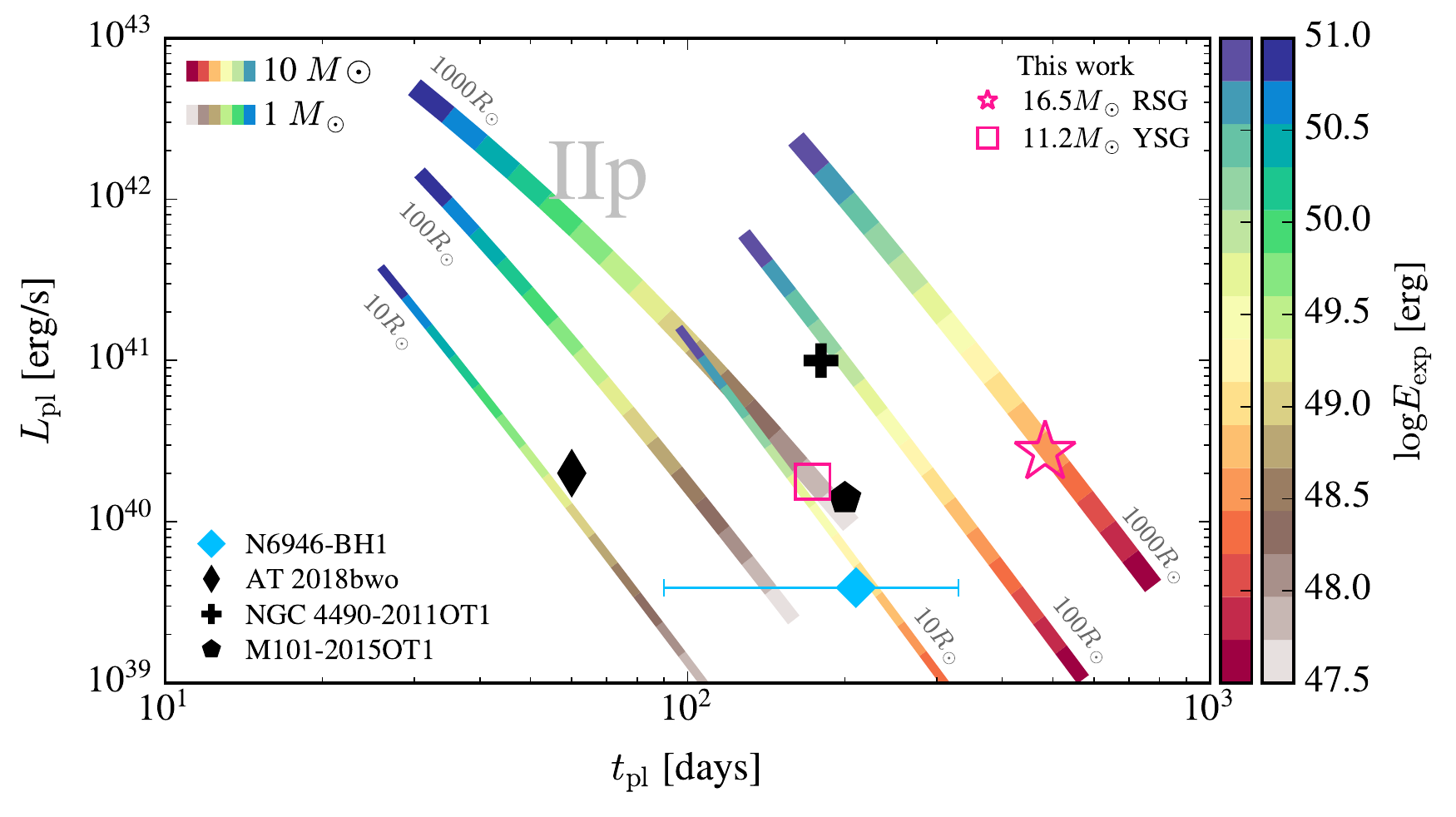}
	\caption{Plateau luminosity versus event duration as a function of explosion energy, ejecta mass and initial radius.  Each shaded curve plots $3 \times 10^{47} < E_{\rm exp} < 10^{51}$ erg for fixed ejecta mass and initial radius.  From thin to thick, the curves correspond to initial radii of 10, 100, and 1000 $\rsun$, respectively.  The two color schemes adopt ejecta masses of 1 and 10 $\msun$ as noted in the legend.  The unfilled points show the expected results for our explosion mechanism when scaled to \mesa models of a $16.5\msun$ RSG and a 11.2$\msun$ YSG.  The filled black points show a sample of luminous red novae (LRNe) with massive, supergiant progenitors (black points; M101-2015OT1: \citealt{2017ApJ...834..107B}, AT 2018bwo: \citealt{2021A&A...653A.134B}, NGC 4490-2011OT1: \citealt{2019A&A...630A..75P}).  The filled blue point is the BH-formation candidate of \citealt{{2017MNRAS.468.4968A}}.  For the LRNe, $t_{\rm pl}$ is the duration of the event from the peak or from the first detection of the outburst to the end of the plateau phase.}
\label{fig:plateau_luminosity_duration}
\end{figure*} 

Fig.~\ref{fig:plateau_luminosity_duration} shows our estimated $L_{\rm pl}$ versus $t_{\rm pl}$ given the ejecta mass, $M_{\rm ej}$, the initial radius of the star, $R_0$, and the explosion energy, $E_{\rm exp}$.  Broadly, these  curves show that explosions of the hydrogen envelopes of RSGs\footnote{We are neglecting, e.g., SN IIn and super-luminous SNe arising from interactions with previously-ejected CSM.} give rise to plateau luminosities from $\sim$few $\times 10^{39}$ erg/s for low-energy explosions up to IIp events with $L_{\rm pl} \gtrsim 10^{41.5}$ erg /s.  Events span plateau durations of many tens of days to a thousand days with the plateau duration increasing for larger $M_{\rm ej}$ and lower $E_{\rm exp}$.  For fixed ejecta mass, larger initial radii of the progenitors also lead to moderately brighter and longer plateaus.  We include in  Fig.~\ref{fig:plateau_luminosity_duration} a large, pink star indicating our estimated simulation results scaled to our $16.5\msun$ \mesa RSG model, for which $M_{\rm ej} = 10.5 \msun$, $R_0 = 840 \rsun$, and $E_{\rm exp} = 4 \times 10^{48}$ erg.  For this model, infall of the convective envelope onto the newborn BH gives rise to a $\sim$few $\times 10^{40}$ erg/s plateau with a duration of $\sim$500 days. For comparison, the unfilled, pink square scales our results to the 11.2$\msun$ YSG model (22$\msun$ at ZAMS) of \citet{2018MNRAS.476.2366F} and \cite{2021ApJ...911....6I}. For this model, $E_{\rm exp} \approx 9.8 \times 10^{47}$ erg, $M_{\rm ej} \approx 1.2\msun$, and $R_0 \approx 690\rsun$.  The resulting plateau luminosity is similar to the RSG model but the plateau duration is shorter by a factor of $\sim$3. 
 
There are many astrophysical phenomena that produce weak explosions of RSGs and give rise to observables that are similar to the mechanism we study in this paper.  The remaining points in Fig.~\ref{fig:plateau_luminosity_duration} plot a sample of observed events.  For explosions of $\sim$few $\times 10^{48}$ erg, the events that are most similar to ours in brightness, duration, ejecta velocity, and color are the possible failed SN N6946-BH1 (filled, blue point with error bars; \citealt{2017MNRAS.468.4968A}) and  the luminous red novae (LRNe; black points) with a more massive RSG or YSG donating the envelope in a common envelope event \citep{2016MNRAS.458..950S,2017ApJ...834..107B,2019A&A...630A..75P,2021A&A...653A.134B,2022ApJ...938....5M}.  We note that many pre-SN outbursts would fall in similar parameter space in Fig.~\ref{fig:plateau_luminosity_duration} (see, e.g., table 1 of \citealt{2022ApJ...936..114M}), but these events have much higher ejecta velocities  and, often, bluer colors than what we expect for our events.  As we discussed in Sec.~\ref{sec:extrap}, the explosion energy and velocity structure of our ejecta remain uncertain as we lack the ability to simulate to smaller radii where more energy is released. If simulating to smaller radii and including magnetic fields increases the explosion energy without shutting off accretion, then our events would fall further up and to the left in Fig.~\ref{fig:plateau_luminosity_duration}, potentially in the regime of low-luminosity or normal SNe IIp. Our mechanism does not produce nickel so could be differentiated by the absence of a radioactive tail in the light curve.

\subsection{Comparison to mass ejection prompted by neutrino emission}
\label{sec:lovegrove}
Previous work has shown that the change in gravitational potential experienced by the RSG envelope due to the emission of $0.1-0.4$ $\msun$ of mass-energy in neutrinos from the core during the proto-neutron star stage can generate a weak shock in the envelope of the star and may eject some mass \citep{1980Ap&SS..69..115N,2013ApJ...769..109L,2018MNRAS.476.2366F,2018MNRAS.477.1225C,2021ApJ...911....6I,2022arXiv220915064S}. 

Using a parameterized treatment of the neutrino radiation, \citet{2018MNRAS.476.2366F} found that their 15$\msun$ RSG ejected its hydrogen envelope with a kinetic energy of $E_{\rm exp}\approx 10^{47}$ erg and an expansion velocity of 70 km/s. They estimate this would result in a plateau luminosity of $2 \times 10^{39}$ erg/s and a duration of $\sim$400 days.  In a follow-up study in which they self-consistently model the core and envelope evolution, \citet{2021ApJ...911....6I} instead found that the same 15$\msun$ RSG progenitor model ejects no unbound mass.  Simulating the proto-neutron star phase of the core until BH-formation with general relativity and neutrino transport is computationally intensive, which limited the number of equations-of-state (EOSs) and pre-SN stars that \citet{2021ApJ...911....6I} could study.  To get around this limitation, \citet{2022arXiv220915064S} instead interpolate tabulated results of core-evolution calculations to inform their hydrodynamical simulations of the envelope's response to the neutrino emission (see also \citealt{Schneider_2020}).  For similar EOS stiffness, their results are consistent with \citet{2021ApJ...911....6I}, but they simulate a much wider range of EOSs.  Across this range they find that for the 15$\msun$ RSG of \citet{2018MNRAS.476.2366F} and \citet{2021ApJ...911....6I}, $E_{\rm exp} \approx 0.3-3 \times 10^{47}$ erg, $M_{\rm ej} \approx 0.6 - 5 \msun$, and $v_{\rm sh} \approx 30 - 100$ km/s.   In most cases, the majority of the hydrogen envelope still accretes, so our mechanism still operates. 

For comparison to our explosion mechanism, we focus on the 15 $\msun$ RSG model presented in fig. 4 of \citet{2018MNRAS.476.2366F} and fig. 6 of \citet{2022arXiv220915064S}.  When comparing their weak shocks launched by the neutrino-mass-loss mechanism to the shock prompted by infall of the convective hydrogen envelope that we study here, we find that the two shocks arrive at the stellar surface at similar times.  This could  result in interesting early-time light curves as shock break-out would carry the imprint of these two different components.  However, the explosion prompted by our mechanism, which unbinds the entire hydrogen envelope in a $\gtrsim$ few $\times$ 10$^{48}$ erg explosion, dominates the energetics and will dominate the light curve at later times (i.e. on the plateau).   

 Mass ejection prompted by the neutrino-emission mechanism has been used to interpret the failed SN candidate N6946-BH1 in which a 25$\msun$ RSG brightened for 3-11 months before dimming by $\sim$ two orders-of-magnitude in the BVRI bands below the progenitor luminosity. The results of our work suggests that collapse of RSGs likely produces explosions of higher energies than previously thought and with plateau luminosities of $\gtrsim$
 few $\times$ 10$^{40}$ erg/s --  brighter than what was seen in N6946-BH1.  Indeed, \citet{2022MNRAS.514.1188B} searched 10 years of ZTF and PTF data for transients whose optical signatures could be similar to the lightcurve of \citet{2013ApJ...769..109L} and found no events. They concluded that such transients are either dimmer than N6946-BH1 or are extremely rare. Available data for N6946-BH1 may be consistent with a dusty merger or outburst in a binary or triple system \citep*[e.g.,][]{2017MNRAS.467.3299K,2022ApJ...937...96M}.   If there is a surviving massive star, however, it should be observable now with JWST whether or not the dusty outflow is spherical or concentrated in an equatorial disk \citep{2023arXiv230511936K}.  For the case of a terminal, weak explosion of a RSG progenitor, radiation hydrodynamical models that include realistic opacities and which follow the expansion of both unbound and bound material over long time scales could help clarify some features of the light curve of N6946-BH1 and its physical origin.

\section{Summary and Conclusions}
\label{sec:conclusion}
If core collapse of a massive RSG or YSG does not lead to the successful explosion of the star in a supernova, a BH is unavoidably formed from the collapsing core. A large fraction of the hydrogen envelope will then fall in towards the BH.  Even in non-rotating RSGs and YSGs, the random velocity field in the convective hydrogen envelope of the star results in finite specific angular momentum at each radius that is larger than that associated with the ISCO of the newly-formed BH \citep{2014MNRAS.439.4011G,2016ApJ...827...40G,Quataert2019,2022MNRAS.511..176A,2022ApJ...929..156G}.  Thus, quiescent, spherical accretion of the infalling hydrogen envelope onto the BH is not possible. Instead, infall of the envelope may generate a luminous transient following an otherwise failed SN. 

In \citet{2022MNRAS.511..176A} (Paper I), we simulated the hydrogen envelope of a non-rotating RSG, modeling a factor of $\sim$20 in radius of the convection zone out to the photosphere of the star.   We found that, prior to collapse, the circularization radius, $\rcirc$, of the material at nearly all radii and all times in the envelope is $\sim 100-2500$ times the radius of the BH ISCO, $\risco$.  In addition, we simulated the initial collapse of the star (down to $r \sim$ 60 $\rcirc$) and found that the specific angular momentum in each shell is roughly conserved as the material falls in. Thus, the values of $\rcirc$ measured prior to collapse are indeed realized during the collapse of the star, motivating us to study the inflow to yet smaller radii where angular momentum becomes dynamically important.  In the present work, we extend the collapse simulations of Paper I to radii $< \rcirc$.  Accretion is facilitated by introduction of a low-pressure sink of radius $\rsink$ at the origin.  Critically, all simulations have $\rsink < \rcirc$ and we resolve the flows at $\sim\rcirc$.  We find that infall of material from the convective envelope results in ejection of the entire hydrogen envelope of the star in an explosion with an energy of at least $\sim10^{48}$ ergs.  

The following summarizes our main results:
\begin{enumerate}
  \item Circularization of  material at small radii drives a shock out through the collapsing star. A substantial fraction of the shocked material has positive total energy, $U$, with $U$ a factor of $\sim$few to 20 times the local escape speed squared (Fig.~\ref {fig:be_ma_timeseries}). Inside the expanding bubble of shocked material, a smaller region of bound gas persists. Through this channel, material flows to yet small radii, deeper into the potential of the BH (Figs.~\ref{fig:be_ma_3planes} and \ref{fig:mach_small}).
  \item Due to the large dispersion in angular momentum of the convective material prior to collapse (Fig.~\ref{fig:hist}), material flows in with a random mean angular momentum vector that changes over time. At small radii, an accretion disk does not form although material is significantly deflected about the instantaneous angular momentum vector. 
   \item The accretion rate is suppressed by a factor of $\gtrsim$ 100 relative to the total mass budget (i.e. relative to if there were no angular momentum and all of the envelope were able to accrete). Accretion is never shut off, however. A small fraction of material liberates energy deep in the potential, resulting in a persistent supply of energy to the outflow (Fig.~\ref{fig:mdot_vs_time}).
  \item The total energy of the star increases monotonically over time, even after the total energy becomes positive and the envelope is unbound.  The total energy at the end of the simulation increases with more physical simulations at higher resolution and with smaller sink radius (Fig.~\ref{fig:bubble_energy}).  Although the shock must climb out through the envelope of the star, the shock velocities are constant due to a roughly constant energy-loading of the explosion. When scaled to a 16.5$\msun$ RSG \mesa model, the shock speeds are $\sim$few $\times$ 100 km/s (Fig.~\ref{fig:shock_radius}). 
  \item  When we extrapolate the results of our long-running simulation (Fig.~\ref{fig:tp_energy}) until the time when the shock reaches the stellar surface, we find that roughly the entire hydrogen envelope is ejected.  When scaling our dimensionless units to a 16.5$\msun$ RSG with a 10.5$\msun$ hydrogen envelope, the explosion has an energy of $E_{\rm exp} \approx 4\times 10^{48}$ erg  and an ejecta velocity of $v_{\rm sh} \approx 200$ km/s (Sec.~\ref{sec:extrap}). For a 11.2$\msun$ YSG with a 1.2$\msun$ hydrogen envelope, $E_{\rm exp} \approx 10^{48}$ erg and $v_{\rm sh} \approx 300$ km/s.
  \item The light curves of explosions generated by our mechanism would be characterized by an initial shock breakout and shock cooling followed by a long plateau as a recombination front recedes through the expanding ejecta. Our reference 16.5$\msun$ RSG model would have a plateau luminosity of $L_{\rm pl} \approx 3 \times 10^{40}$ erg/s and a duration of $\approx 500$ days; the YSG model would have similar $L_{\rm pl}$ but a shorter duration of $\approx$ 200 days (Fig.~\ref{fig:plateau_luminosity_duration}).  
\end{enumerate}

The explosions generated by our mechanism are quite similar to luminous red novae (LRNe) not only in luminosity and duration (Fig.~\ref{fig:plateau_luminosity_duration}), but also in ejecta velocities and in qualitative features of the characteristic double-peaked light curves of LRNe with an initial blue peak (due to shock breakout/cooling lasting $\sim$ few to tens of days in our events) followed by a red plateau of hundreds of days \citep{2023ApJ...948..137K}. It is thus possible that some LRNe and transients in that part of color-duration-luminosity space are, in fact, signatures of BH-formation in stellar core-collapse. Indeed, the failed SN candidate N6946-BH1 is also similar to LRNe.  In our models, however, if the progenitor of N6946-BH1 is a 25$\msun$ RSG, we would expect a somewhat higher explosion energy and luminosity from infall of the convective envelope following the collapse to form a BH.   Further radiation hydrodynamical modeling of weak explosions of supergiants could help shed light on the physical origin of N6946-BH1. It would also be valuable to carry out collapse and explosion calculations like those done here using as realistic as possible pre-collapse RSG structure. Although the polytropic models we have used in our work are reasonable approximations to RSG structure, the importance of understanding the origin of N6946-BH1 motivates a more detailed study targeted at this event.

Our estimates for the explosion energy of the hydrogen envelope are likely lower limits. We have shown that the explosion energy and ejecta velocity increase with increasing resolution and decreasing $\rsink$ (Figs.~\ref {fig:bubble_energy} and \ref {fig:shock_radius}). Yet higher resolution and yet smaller $\rsink$ would likely increase the energy further, though it is unclear how close we are to converging in these pure hydrodynamical simulations.  We also do not run the simulation until the true end of the event (that is, when accretion terminates).  In Sec.~\ref{sec:extrap}, we extrapolate the explosion energy to the time when the (non-spherical) shock reaches the surface of the star, but there are channels through which inflowing, still bound material will continue to fall in to small radii for at least the full dynamical time of the envelope.  The most important influence on the energy and morphology of the explosion is probably that we do not include magnetic fields in our simulations. Future simulations that include magnetic fields will help us understand the range of explosion energies that may arise from BH formation in RSGs.  If the true explosion energies are much higher than the lower limits we give in this work, then the luminosities and durations could be similar to low-luminosity or even normal SN IIps, though without a radioactive tail. 

\section*{Acknowledgements}
This research was conducted on Ohlone\footnote{https://sogoreate-landtrust.org/} and Lenni-Lenape\footnote{https://nlltribe.com/} lands.  The authors wish to thank Sean Ressler for providing a sample adaptive mesh refinement criteria for {\sc athena++} and both Drummond Fielding and Sean Ressler for their helpful suggestions about implementing sinks in subsonic flows. We are also grateful to Lars Bildsten, Tamar Faran, Wynn Jacobson-Gal{\'a}n, Jared Goldberg, Yan-Fei Jiang, Raffaella Margutti, Morgan MacLeod, Kishore Patra, Sophie L. Schr\o der, and Benny T.-H. Tsang for helpful discussions. 

A.A. gratefully acknowledges support from the University of California, Berkeley Fellowship, the Cranor Fellowship at U.C. Berkeley, the National Science Foundation Graduate Research Fellowship under Grant No. DGE 1752814, and the  Gordon and Betty Moore Foundation through Grant GBMF5076.  E.Q. was supported in part by a Simons Investigator award from the Simons Foundation. 
This research benefited from interactions with Jim Fuller, Viraj Karambelkar, Wenbin Lu, and Ben Margalit at workshops funded by the Gordon and Betty Moore Foundation through Grant GBMF5076.

The simulations presented in this article were performed on computational resources managed and supported by Princeton Research Computing, a consortium of groups including the Princeton Institute for Computational Science and Engineering (PICSciE) and the Office of Information Technology's High Performance Computing Center and Visualization Laboratory at Princeton University. This project was made possible by the following publicly available software: {\sc astropy} \citep{astropy:2013, astropy:2018}, {\sc athena++} \citep{2020ApJS..249....4S}, {\sc matplotlib} \citep{Hunter:2007}, {\sc mesa} \citep{2011ApJS..192....3P,2013ApJS..208....4P,2015ApJS..220...15P,2018ApJS..234...34P,2019ApJS..243...10P},  {\sc mesa sdk} \citep{richard_townsend_2020_3706650},  {\sc numpy} \citep{harris2020array}, {\sc yt} \citep{2011ApJS..192....9T}.


\section*{Data Availability}
Data related to the results of simulations in this article will be shared on reasonable request to the corresponding author via e-mail.



\bibliographystyle{mnras}
\bibliography{references} 

\begin{thebibliography}{}
\makeatletter
\relax
\def\mn@urlcharsother{\let\do\@makeother \do\$\do\&\do\#\do\^\do\_\do\%\do\~}
\def\mn@doi{\begingroup\mn@urlcharsother \@ifnextchar [ {\mn@doi@}
  {\mn@doi@[]}}
\def\mn@doi@[#1]#2{\def\@tempa{#1}\ifx\@tempa\@empty \href
  {http://dx.doi.org/#2} {doi:#2}\else \href {http://dx.doi.org/#2} {#1}\fi
  \endgroup}
\def\mn@eprint#1#2{\mn@eprint@#1:#2::\@nil}
\def\mn@eprint@arXiv#1{\href {http://arxiv.org/abs/#1} {{\tt arXiv:#1}}}
\def\mn@eprint@dblp#1{\href {http://dblp.uni-trier.de/rec/bibtex/#1.xml}
  {dblp:#1}}
\def\mn@eprint@#1:#2:#3:#4\@nil{\def\@tempa {#1}\def\@tempb {#2}\def\@tempc
  {#3}\ifx \@tempc \@empty \let \@tempc \@tempb \let \@tempb \@tempa \fi \ifx
  \@tempb \@empty \def\@tempb {arXiv}\fi \@ifundefined
  {mn@eprint@\@tempb}{\@tempb:\@tempc}{\expandafter \expandafter \csname
  mn@eprint@\@tempb\endcsname \expandafter{\@tempc}}}

\bibitem[\protect\citeauthoryear{{Adams}, {Kochanek}, {Gerke}, {Stanek}  \&
  {Dai}}{{Adams} et~al.}{2017}]{2017MNRAS.468.4968A}
{Adams} S.~M.,  {Kochanek} C.~S.,  {Gerke} J.~R.,  {Stanek} K.~Z.,   {Dai} X.,
  2017, \mn@doi [\mnras] {10.1093/mnras/stx816}, \href
  {https://ui.adsabs.harvard.edu/abs/2017MNRAS.468.4968A} {468, 4968}

\bibitem[\protect\citeauthoryear{{Antoni} \& {Quataert}}{{Antoni} \&
  {Quataert}}{2022}]{2022MNRAS.511..176A}
{Antoni} A.,  {Quataert} E.,  2022, \mn@doi [\mnras] {10.1093/mnras/stab3776},
  \href {https://ui.adsabs.harvard.edu/abs/2022MNRAS.511..176A} {511, 176}

\bibitem[\protect\citeauthoryear{{Astropy Collaboration} et~al.,}{{Astropy
  Collaboration} et~al.}{2013}]{astropy:2013}
{Astropy Collaboration} et~al., 2013, \mn@doi [\aap]
  {10.1051/0004-6361/201322068}, \href
  {http://adsabs.harvard.edu/abs/2013A%26A...558A..33A} {558, A33}

\bibitem[\protect\citeauthoryear{{Astropy Collaboration} et~al.,}{{Astropy
  Collaboration} et~al.}{2018}]{astropy:2018}
{Astropy Collaboration} et~al., 2018, \mn@doi [aj] {10.3847/1538-3881/aabc4f},
  \href {https://ui.adsabs.harvard.edu/abs/2018AJ....156..123A} {156, 123}

\bibitem[\protect\citeauthoryear{{Batta} \& {Ramirez-Ruiz}}{{Batta} \&
  {Ramirez-Ruiz}}{2019}]{2019arXiv190404835B}
{Batta} A.,  {Ramirez-Ruiz} E.,  2019, arXiv e-prints, \href
  {https://ui.adsabs.harvard.edu/abs/2019arXiv190404835B} {p. arXiv:1904.04835}

\bibitem[\protect\citeauthoryear{{Blagorodnova} et~al.,}{{Blagorodnova}
  et~al.}{2017}]{2017ApJ...834..107B}
{Blagorodnova} N.,  et~al., 2017, \mn@doi [\apj] {10.3847/1538-4357/834/2/107},
  \href {https://ui.adsabs.harvard.edu/abs/2017ApJ...834..107B} {834, 107}

\bibitem[\protect\citeauthoryear{{Blagorodnova} et~al.,}{{Blagorodnova}
  et~al.}{2021}]{2021A&A...653A.134B}
{Blagorodnova} N.,  et~al., 2021, \mn@doi [\aap] {10.1051/0004-6361/202140525},
  \href {https://ui.adsabs.harvard.edu/abs/2021A&A...653A.134B} {653, A134}

\bibitem[\protect\citeauthoryear{{Boccioli}, {Roberti}, {Limongi}, {Mathews}
  \& {Chieffi}}{{Boccioli} et~al.}{2023}]{2023ApJ...949...17B}
{Boccioli} L.,  {Roberti} L.,  {Limongi} M.,  {Mathews} G.~J.,   {Chieffi} A.,
  2023, \mn@doi [\apj] {10.3847/1538-4357/acc06a}, \href
  {https://ui.adsabs.harvard.edu/abs/2023ApJ...949...17B} {949, 17}

\bibitem[\protect\citeauthoryear{{Burrows}, {Radice}, {Vartanyan}, {Nagakura},
  {Skinner}  \& {Dolence}}{{Burrows} et~al.}{2020}]{2020MNRAS.491.2715B}
{Burrows} A.,  {Radice} D.,  {Vartanyan} D.,  {Nagakura} H.,  {Skinner} M.~A.,
   {Dolence} J.~C.,  2020, \mn@doi [\mnras] {10.1093/mnras/stz3223}, \href
  {https://ui.adsabs.harvard.edu/abs/2020MNRAS.491.2715B} {491, 2715}

\bibitem[\protect\citeauthoryear{{Byrne} \& {Fraser}}{{Byrne} \&
  {Fraser}}{2022}]{2022MNRAS.514.1188B}
{Byrne} R.~A.,  {Fraser} M.,  2022, \mn@doi [\mnras] {10.1093/mnras/stac1308},
  \href {https://ui.adsabs.harvard.edu/abs/2022MNRAS.514.1188B} {514, 1188}

\bibitem[\protect\citeauthoryear{{Couch}, {Warren}  \& {O'Connor}}{{Couch}
  et~al.}{2020}]{2020ApJ...890..127C}
{Couch} S.~M.,  {Warren} M.~L.,   {O'Connor} E.~P.,  2020, \mn@doi [\apj]
  {10.3847/1538-4357/ab609e}, \href
  {https://ui.adsabs.harvard.edu/abs/2020ApJ...890..127C} {890, 127}

\bibitem[\protect\citeauthoryear{{Coughlin}, {Quataert}, {Fern{\'a}ndez}  \&
  {Kasen}}{{Coughlin} et~al.}{2018}]{2018MNRAS.477.1225C}
{Coughlin} E.~R.,  {Quataert} E.,  {Fern{\'a}ndez} R.,   {Kasen} D.,  2018,
  \mn@doi [\mnras] {10.1093/mnras/sty667}, \href
  {https://ui.adsabs.harvard.edu/abs/2018MNRAS.477.1225C} {477, 1225}

\bibitem[\protect\citeauthoryear{{Dessart}, {Livne}  \& {Waldman}}{{Dessart}
  et~al.}{2010}]{2010MNRAS.405.2113D}
{Dessart} L.,  {Livne} E.,   {Waldman} R.,  2010, \mn@doi [\mnras]
  {10.1111/j.1365-2966.2010.16626.x}, \href
  {https://ui.adsabs.harvard.edu/abs/2010MNRAS.405.2113D} {405, 2113}

\bibitem[\protect\citeauthoryear{{Dexter} \& {Kasen}}{{Dexter} \&
  {Kasen}}{2013}]{DK2013}
{Dexter} J.,  {Kasen} D.,  2013, \mn@doi [\apj] {10.1088/0004-637X/772/1/30},
  \href {https://ui.adsabs.harvard.edu/abs/2013ApJ...772...30D} {772, 30}

\bibitem[\protect\citeauthoryear{{Fern{\'a}ndez}, {Quataert}, {Kashiyama}  \&
  {Coughlin}}{{Fern{\'a}ndez} et~al.}{2018}]{2018MNRAS.476.2366F}
{Fern{\'a}ndez} R.,  {Quataert} E.,  {Kashiyama} K.,   {Coughlin} E.~R.,  2018,
  \mn@doi [\mnras] {10.1093/mnras/sty306}, \href
  {https://ui.adsabs.harvard.edu/abs/2018MNRAS.476.2366F} {476, 2366}

\bibitem[\protect\citeauthoryear{{Fuller} \& {Ma}}{{Fuller} \&
  {Ma}}{2019}]{2019ApJ...881L...1F}
{Fuller} J.,  {Ma} L.,  2019, \mn@doi [\apjl] {10.3847/2041-8213/ab339b}, \href
  {https://ui.adsabs.harvard.edu/abs/2019ApJ...881L...1F} {881, L1}

\bibitem[\protect\citeauthoryear{{Gilkis} \& {Soker}}{{Gilkis} \&
  {Soker}}{2014}]{2014MNRAS.439.4011G}
{Gilkis} A.,  {Soker} N.,  2014, \mn@doi [\mnras] {10.1093/mnras/stu257}, \href
  {https://ui.adsabs.harvard.edu/abs/2014MNRAS.439.4011G} {439, 4011}

\bibitem[\protect\citeauthoryear{{Gilkis} \& {Soker}}{{Gilkis} \&
  {Soker}}{2016}]{2016ApJ...827...40G}
{Gilkis} A.,  {Soker} N.,  2016, \mn@doi [\apj] {10.3847/0004-637X/827/1/40},
  \href {https://ui.adsabs.harvard.edu/abs/2016ApJ...827...40G} {827, 40}

\bibitem[\protect\citeauthoryear{{Goldberg}, {Bildsten}  \&
  {Paxton}}{{Goldberg} et~al.}{2019}]{Goldberg2019}
{Goldberg} J.~A.,  {Bildsten} L.,   {Paxton} B.,  2019, \mn@doi [\apj]
  {10.3847/1538-4357/ab22b6}, \href
  {https://ui.adsabs.harvard.edu/abs/2019ApJ...879....3G} {879, 3}

\bibitem[\protect\citeauthoryear{{Goldberg}, {Jiang}  \& {Bildsten}}{{Goldberg}
  et~al.}{2022}]{2022ApJ...929..156G}
{Goldberg} J.~A.,  {Jiang} Y.-F.,   {Bildsten} L.,  2022, \mn@doi [\apj]
  {10.3847/1538-4357/ac5ab3}, \href
  {https://ui.adsabs.harvard.edu/abs/2022ApJ...929..156G} {929, 156}

\bibitem[\protect\citeauthoryear{{Gottlieb}, {Lalakos}, {Bromberg}, {Liska}  \&
  {Tchekhovskoy}}{{Gottlieb} et~al.}{2022}]{2022MNRAS.510.4962G}
{Gottlieb} O.,  {Lalakos} A.,  {Bromberg} O.,  {Liska} M.,   {Tchekhovskoy} A.,
   2022, \mn@doi [\mnras] {10.1093/mnras/stab3784}, \href
  {https://ui.adsabs.harvard.edu/abs/2022MNRAS.510.4962G} {510, 4962}

\bibitem[\protect\citeauthoryear{Harris et~al.,}{Harris
  et~al.}{2020}]{harris2020array}
Harris C.~R.,  et~al., 2020, \mn@doi [Nature] {10.1038/s41586-020-2649-2}, 585,
  357

\bibitem[\protect\citeauthoryear{{Hu}, {Inayoshi}, {Haiman}, {Quataert}  \&
  {Kuiper}}{{Hu} et~al.}{2022}]{2022ApJ...934..132H}
{Hu} H.,  {Inayoshi} K.,  {Haiman} Z.,  {Quataert} E.,   {Kuiper} R.,  2022,
  \mn@doi [\apj] {10.3847/1538-4357/ac75d8}, \href
  {https://ui.adsabs.harvard.edu/abs/2022ApJ...934..132H} {934, 132}

\bibitem[\protect\citeauthoryear{Hunter}{Hunter}{2007}]{Hunter:2007}
Hunter J.~D.,  2007, \mn@doi [Computing in Science \& Engineering]
  {10.1109/MCSE.2007.55}, 9, 90

\bibitem[\protect\citeauthoryear{{Ivanov} \& {Fern{\'a}ndez}}{{Ivanov} \&
  {Fern{\'a}ndez}}{2021}]{2021ApJ...911....6I}
{Ivanov} M.,  {Fern{\'a}ndez} R.,  2021, \mn@doi [\apj]
  {10.3847/1538-4357/abe59e}, \href
  {https://ui.adsabs.harvard.edu/abs/2021ApJ...911....6I} {911, 6}

\bibitem[\protect\citeauthoryear{{Karambelkar} et~al.,}{{Karambelkar}
  et~al.}{2023}]{2023ApJ...948..137K}
{Karambelkar} V.~R.,  et~al., 2023, \mn@doi [\apj] {10.3847/1538-4357/acc2b9},
  \href {https://ui.adsabs.harvard.edu/abs/2023ApJ...948..137K} {948, 137}

\bibitem[\protect\citeauthoryear{{Kashi} \& {Soker}}{{Kashi} \&
  {Soker}}{2017}]{2017MNRAS.467.3299K}
{Kashi} A.,  {Soker} N.,  2017, \mn@doi [\mnras] {10.1093/mnras/stx240}, \href
  {https://ui.adsabs.harvard.edu/abs/2017MNRAS.467.3299K} {467, 3299}

\bibitem[\protect\citeauthoryear{{Kochanek}}{{Kochanek}}{2023}]{2023arXiv230511936K}
{Kochanek} C.~S.,  2023, \mn@doi [arXiv e-prints] {10.48550/arXiv.2305.11936},
  \href {https://ui.adsabs.harvard.edu/abs/2023arXiv230511936K} {p.
  arXiv:2305.11936}

\bibitem[\protect\citeauthoryear{{Kochanek}, {Beacom}, {Kistler}, {Prieto},
  {Stanek}, {Thompson}  \& {Y{\"u}ksel}}{{Kochanek}
  et~al.}{2008}]{2008ApJ...684.1336K}
{Kochanek} C.~S.,  {Beacom} J.~F.,  {Kistler} M.~D.,  {Prieto} J.~L.,  {Stanek}
  K.~Z.,  {Thompson} T.~A.,   {Y{\"u}ksel} H.,  2008, \mn@doi [\apj]
  {10.1086/590053}, \href
  {https://ui.adsabs.harvard.edu/abs/2008ApJ...684.1336K} {684, 1336}

\bibitem[\protect\citeauthoryear{{Lovegrove} \& {Woosley}}{{Lovegrove} \&
  {Woosley}}{2013}]{2013ApJ...769..109L}
{Lovegrove} E.,  {Woosley} S.~E.,  2013, \mn@doi [\apj]
  {10.1088/0004-637X/769/2/109}, \href
  {https://ui.adsabs.harvard.edu/abs/2013ApJ...769..109L} {769, 109}

\bibitem[\protect\citeauthoryear{{MacFadyen} \& {Woosley}}{{MacFadyen} \&
  {Woosley}}{1999}]{1999ApJ...524..262M}
{MacFadyen} A.~I.,  {Woosley} S.~E.,  1999, \mn@doi [\apj] {10.1086/307790},
  \href {https://ui.adsabs.harvard.edu/abs/1999ApJ...524..262M} {524, 262}

\bibitem[\protect\citeauthoryear{{MacLeod}, {De}  \& {Loeb}}{{MacLeod}
  et~al.}{2022}]{2022ApJ...937...96M}
{MacLeod} M.,  {De} K.,   {Loeb} A.,  2022, \mn@doi [\apj]
  {10.3847/1538-4357/ac8c31}, \href
  {https://ui.adsabs.harvard.edu/abs/2022ApJ...937...96M} {937, 96}

\bibitem[\protect\citeauthoryear{{Matsumoto} \& {Metzger}}{{Matsumoto} \&
  {Metzger}}{2022a}]{2022ApJ...936..114M}
{Matsumoto} T.,  {Metzger} B.~D.,  2022a, \mn@doi [\apj]
  {10.3847/1538-4357/ac892c}, \href
  {https://ui.adsabs.harvard.edu/abs/2022ApJ...936..114M} {936, 114}

\bibitem[\protect\citeauthoryear{{Matsumoto} \& {Metzger}}{{Matsumoto} \&
  {Metzger}}{2022b}]{2022ApJ...938....5M}
{Matsumoto} T.,  {Metzger} B.~D.,  2022b, \mn@doi [\apj]
  {10.3847/1538-4357/ac6269}, \href
  {https://ui.adsabs.harvard.edu/abs/2022ApJ...938....5M} {938, 5}

\bibitem[\protect\citeauthoryear{{Murguia-Berthier}, {Batta}, {Janiuk},
  {Ramirez-Ruiz}, {Mandel}, {Noble}  \& {Everson}}{{Murguia-Berthier}
  et~al.}{2020}]{2020ApJ...901L..24M}
{Murguia-Berthier} A.,  {Batta} A.,  {Janiuk} A.,  {Ramirez-Ruiz} E.,  {Mandel}
  I.,  {Noble} S.~C.,   {Everson} R.~W.,  2020, \mn@doi [\apjl]
  {10.3847/2041-8213/abb818}, \href
  {https://ui.adsabs.harvard.edu/abs/2020ApJ...901L..24M} {901, L24}

\bibitem[\protect\citeauthoryear{{Nadyozhin}}{{Nadyozhin}}{1980}]{1980Ap&SS..69..115N}
{Nadyozhin} D.~K.,  1980, \mn@doi [\apss] {10.1007/BF00638971}, \href
  {https://ui.adsabs.harvard.edu/abs/1980Ap&SS..69..115N} {69, 115}

\bibitem[\protect\citeauthoryear{{Neustadt}, {Kochanek}, {Stanek}, {Basinger},
  {Jayasinghe}, {Garling}, {Adams}  \& {Gerke}}{{Neustadt}
  et~al.}{2021}]{2021MNRAS.508..516N}
{Neustadt} J.~M.~M.,  {Kochanek} C.~S.,  {Stanek} K.~Z.,  {Basinger} C.,
  {Jayasinghe} T.,  {Garling} C.~T.,  {Adams} S.~M.,   {Gerke} J.,  2021,
  \mn@doi [\mnras] {10.1093/mnras/stab2605}, \href
  {https://ui.adsabs.harvard.edu/abs/2021MNRAS.508..516N} {508, 516}

\bibitem[\protect\citeauthoryear{{Ohsuga}, {Mori}, {Nakamoto}  \&
  {Mineshige}}{{Ohsuga} et~al.}{2005}]{2005ApJ...628..368O}
{Ohsuga} K.,  {Mori} M.,  {Nakamoto} T.,   {Mineshige} S.,  2005, \mn@doi
  [\apj] {10.1086/430728}, \href
  {https://ui.adsabs.harvard.edu/abs/2005ApJ...628..368O} {628, 368}

\bibitem[\protect\citeauthoryear{{Pang}, {Pen}, {Matzner}, {Green}  \&
  {Liebend{\"o}rfer}}{{Pang} et~al.}{2011}]{2011MNRAS.415.1228P}
{Pang} B.,  {Pen} U.-L.,  {Matzner} C.~D.,  {Green} S.~R.,   {Liebend{\"o}rfer}
  M.,  2011, \mn@doi [\mnras] {10.1111/j.1365-2966.2011.18748.x}, \href
  {https://ui.adsabs.harvard.edu/abs/2011MNRAS.415.1228P} {415, 1228}

\bibitem[\protect\citeauthoryear{{Papish} \& {Soker}}{{Papish} \&
  {Soker}}{2011}]{2011MNRAS.416.1697P}
{Papish} O.,  {Soker} N.,  2011, \mn@doi [\mnras]
  {10.1111/j.1365-2966.2011.18671.x}, \href
  {https://ui.adsabs.harvard.edu/abs/2011MNRAS.416.1697P} {416, 1697}

\bibitem[\protect\citeauthoryear{{Pastorello} et~al.,}{{Pastorello}
  et~al.}{2019}]{2019A&A...630A..75P}
{Pastorello} A.,  et~al., 2019, \mn@doi [\aap] {10.1051/0004-6361/201935999},
  \href {https://ui.adsabs.harvard.edu/abs/2019A&A...630A..75P} {630, A75}

\bibitem[\protect\citeauthoryear{{Paxton}, {Bildsten}, {Dotter}, {Herwig},
  {Lesaffre}  \& {Timmes}}{{Paxton} et~al.}{2011}]{2011ApJS..192....3P}
{Paxton} B.,  {Bildsten} L.,  {Dotter} A.,  {Herwig} F.,  {Lesaffre} P.,
  {Timmes} F.,  2011, \mn@doi [\apjs] {10.1088/0067-0049/192/1/3}, \href
  {https://ui.adsabs.harvard.edu/abs/2011ApJS..192....3P} {192, 3}

\bibitem[\protect\citeauthoryear{{Paxton} et~al.,}{{Paxton}
  et~al.}{2013}]{2013ApJS..208....4P}
{Paxton} B.,  et~al., 2013, \mn@doi [\apjs] {10.1088/0067-0049/208/1/4}, \href
  {https://ui.adsabs.harvard.edu/abs/2013ApJS..208....4P} {208, 4}

\bibitem[\protect\citeauthoryear{{Paxton} et~al.,}{{Paxton}
  et~al.}{2015}]{2015ApJS..220...15P}
{Paxton} B.,  et~al., 2015, \mn@doi [\apjs] {10.1088/0067-0049/220/1/15}, \href
  {https://ui.adsabs.harvard.edu/abs/2015ApJS..220...15P} {220, 15}

\bibitem[\protect\citeauthoryear{{Paxton} et~al.,}{{Paxton}
  et~al.}{2018}]{2018ApJS..234...34P}
{Paxton} B.,  et~al., 2018, \mn@doi [\apjs] {10.3847/1538-4365/aaa5a8}, \href
  {https://ui.adsabs.harvard.edu/abs/2018ApJS..234...34P} {234, 34}

\bibitem[\protect\citeauthoryear{{Paxton} et~al.,}{{Paxton}
  et~al.}{2019}]{2019ApJS..243...10P}
{Paxton} B.,  et~al., 2019, \mn@doi [\apjs] {10.3847/1538-4365/ab2241}, \href
  {https://ui.adsabs.harvard.edu/abs/2019ApJS..243...10P} {243, 10}

\bibitem[\protect\citeauthoryear{{Perna}, {Lazzati}  \& {Cantiello}}{{Perna}
  et~al.}{2018}]{2018ApJ...859...48P}
{Perna} R.,  {Lazzati} D.,   {Cantiello} M.,  2018, \mn@doi [\apj]
  {10.3847/1538-4357/aabcc1}, \href
  {https://ui.adsabs.harvard.edu/abs/2018ApJ...859...48P} {859, 48}

\bibitem[\protect\citeauthoryear{{Popov}}{{Popov}}{1993}]{Popov1993}
{Popov} D.~V.,  1993, \mn@doi [\apj] {10.1086/173117}, \href
  {https://ui.adsabs.harvard.edu/abs/1993ApJ...414..712P} {414, 712}

\bibitem[\protect\citeauthoryear{{Quataert} \& {Kasen}}{{Quataert} \&
  {Kasen}}{2012}]{2012MNRAS.419L...1Q}
{Quataert} E.,  {Kasen} D.,  2012, \mn@doi [\mnras]
  {10.1111/j.1745-3933.2011.01151.x}, \href
  {https://ui.adsabs.harvard.edu/abs/2012MNRAS.419L...1Q} {419, L1}

\bibitem[\protect\citeauthoryear{{Quataert}, {Lecoanet}  \&
  {Coughlin}}{{Quataert} et~al.}{2019}]{Quataert2019}
{Quataert} E.,  {Lecoanet} D.,   {Coughlin} E.~R.,  2019, \mn@doi [\mnras]
  {10.1093/mnrasl/slz031}, \href
  {https://ui.adsabs.harvard.edu/abs/2019MNRAS.485L..83Q} {485, L83}

\bibitem[\protect\citeauthoryear{{Ressler}, {Quataert}  \& {Stone}}{{Ressler}
  et~al.}{2020}]{2020MNRAS.492.3272R}
{Ressler} S.~M.,  {Quataert} E.,   {Stone} J.~M.,  2020, \mn@doi [\mnras]
  {10.1093/mnras/stz3605}, \href
  {https://ui.adsabs.harvard.edu/abs/2020MNRAS.492.3272R} {492, 3272}

\bibitem[\protect\citeauthoryear{{Schneider} \& {O'Connor}}{{Schneider} \&
  {O'Connor}}{2022}]{2022arXiv220915064S}
{Schneider} A.~S.,  {O'Connor} E.,  2022, arXiv e-prints, \href
  {https://ui.adsabs.harvard.edu/abs/2022arXiv220915064S} {p. arXiv:2209.15064}

\bibitem[\protect\citeauthoryear{{Smith} et~al.,}{{Smith}
  et~al.}{2016}]{2016MNRAS.458..950S}
{Smith} N.,  et~al., 2016, \mn@doi [\mnras] {10.1093/mnras/stw219}, \href
  {https://ui.adsabs.harvard.edu/abs/2016MNRAS.458..950S} {458, 950}

\bibitem[\protect\citeauthoryear{{Soker}}{{Soker}}{2010}]{2010MNRAS.401.2793S}
{Soker} N.,  2010, \mn@doi [\mnras] {10.1111/j.1365-2966.2009.15862.x}, \href
  {https://ui.adsabs.harvard.edu/abs/2010MNRAS.401.2793S} {401, 2793}

\bibitem[\protect\citeauthoryear{{Stone}, {Tomida}, {White}  \&
  {Felker}}{{Stone} et~al.}{2020}]{2020ApJS..249....4S}
{Stone} J.~M.,  {Tomida} K.,  {White} C.~J.,   {Felker} K.~G.,  2020, \mn@doi
  [\apjs] {10.3847/1538-4365/ab929b}, \href
  {https://ui.adsabs.harvard.edu/abs/2020ApJS..249....4S} {249, 4}

\bibitem[\protect\citeauthoryear{{Sukhbold}, {Ertl}, {Woosley}, {Brown}  \&
  {Janka}}{{Sukhbold} et~al.}{2016}]{2016ApJ...821...38S}
{Sukhbold} T.,  {Ertl} T.,  {Woosley} S.~E.,  {Brown} J.~M.,   {Janka} H.~T.,
  2016, \mn@doi [\apj] {10.3847/0004-637X/821/1/38}, \href
  {https://ui.adsabs.harvard.edu/abs/2016ApJ...821...38S} {821, 38}

\bibitem[\protect\citeauthoryear{Townsend}{Townsend}{2020}]{richard_townsend_2020_3706650}
Townsend R.,  2020, MESA SDK for Linux, \mn@doi{10.5281/zenodo.3706650}, \url
  {https://doi.org/10.5281/zenodo.3706650}

\bibitem[\protect\citeauthoryear{{Turk}, {Smith}, {Oishi}, {Skory}, {Skillman},
  {Abel}  \& {Norman}}{{Turk} et~al.}{2011}]{2011ApJS..192....9T}
{Turk} M.~J.,  {Smith} B.~D.,  {Oishi} J.~S.,  {Skory} S.,  {Skillman} S.~W.,
  {Abel} T.,   {Norman} M.~L.,  2011, \mn@doi [The Astrophysical Journal
  Supplement Series] {10.1088/0067-0049/192/1/9}, \href
  {https://ui.adsabs.harvard.edu/abs/2011ApJS..192....9T} {192, 9}

\bibitem[\protect\citeauthoryear{{Ugliano}, {Janka}, {Marek}  \&
  {Arcones}}{{Ugliano} et~al.}{2012}]{2012ApJ...757...69U}
{Ugliano} M.,  {Janka} H.-T.,  {Marek} A.,   {Arcones} A.,  2012, \mn@doi
  [\apj] {10.1088/0004-637X/757/1/69}, \href
  {https://ui.adsabs.harvard.edu/abs/2012ApJ...757...69U} {757, 69}

\bibitem[\protect\citeauthoryear{{Woosley}}{{Woosley}}{1993}]{1993ApJ...405..273W}
{Woosley} S.~E.,  1993, \mn@doi [\apj] {10.1086/172359}, \href
  {https://ui.adsabs.harvard.edu/abs/1993ApJ...405..273W} {405, 273}

\bibitem[\protect\citeauthoryear{{Woosley} \& {Heger}}{{Woosley} \&
  {Heger}}{2012}]{2012ApJ...752...32W}
{Woosley} S.~E.,  {Heger} A.,  2012, \mn@doi [\apj]
  {10.1088/0004-637X/752/1/32}, \href
  {https://ui.adsabs.harvard.edu/abs/2012ApJ...752...32W} {752, 32}

\bibitem[\protect\citeauthoryear{{Zhang}, {Woosley}  \& {Heger}}{{Zhang}
  et~al.}{2008}]{2008ApJ...679..639Z}
{Zhang} W.,  {Woosley} S.~E.,   {Heger} A.,  2008, \mn@doi [\apj]
  {10.1086/526404}, \href
  {https://ui.adsabs.harvard.edu/abs/2008ApJ...679..639Z} {679, 639}

\bibitem[\protect\citeauthoryear{da Silva~Schneider, O'Connor, Granqvist,
  Betranhandy  \& Couch}{da~Silva~Schneider et~al.}{2020}]{Schneider_2020}
da Silva~Schneider A.,  O'Connor E.,  Granqvist E.,  Betranhandy A.,   Couch
  S.~M.,  2020, \mn@doi [The Astrophysical Journal] {10.3847/1538-4357/ab8308},
  894, 4

\makeatother
\end{thebibliography}




\bsp	
\label{lastpage}
\end{document}